\documentclass[]{aa}
\usepackage{amsmath,amssymb}
\usepackage{graphics,latexsym,epsfig,txfonts,graphicx,natbib}
\usepackage{url}

\bibpunct{(}{)}{;}{a}{}{,}

\newcommand{\beqn}{\begin{equation}}
\newcommand{\eeqn}{\end{equation}}

\newcommand{\dd}{{\rm d}}

\newcommand{\secref}[1]{Sec.~\ref{#1}}

\defcitealias{schrabbaACS}{S07}

\begin{document}

\title{The non-Gaussianity of the cosmic shear likelihood\\ or \\ How odd is the \textit{Chandra} Deep Field South?}

\author{J.\ Hartlap\inst{1} \and T. Schrabback\inst{2,1} \and P. Simon\inst{3}  \and P.\ Schneider\inst{1}}

\institute{
	Argelander-Institut f\"ur Astronomie, Universit\"at Bonn, Auf dem H\"ugel 71, D-53121 Bonn, Germany \and Leiden Observatory, Universiteit Leiden, Niels Bohrweg 2, NL-2333 CA Leiden, The Netherlands \and The Scottish Universities Physics Alliance (SUPA), Institute for Astronomy, School of Physics, \\University of Edinburgh, Royal Observatory, Blackford Hill, Edinburgh EH9 3HJ, UK}

\date{Received 22 January 2009 / Accepted 19 June 2009}
\authorrunning{Hartlap et al.}
\titlerunning{The non-Gaussianity of the cosmic shear likelihood}

\abstract{}{We study the validity of the approximation of a Gaussian cosmic shear likelihood. We estimate the true likelihood for a fiducial cosmological model from a large set of ray-tracing simulations and investigate the impact of non-Gaussianity on cosmological parameter estimation.
We investigate how odd the recently reported very low value of $\sigma_8$ really is as derived from the \textit{Chandra} Deep Field South (CDFS) using cosmic shear by taking the non-Gaussianity of the likelihood into account as well as the possibility of biases coming from the way the CDFS was selected.}
{A brute force approach to estimating the likelihood from simulations must fail because of the high dimensionality of the problem. We therefore use independent component analysis to transform the cosmic shear correlation functions to a new basis, in which the likelihood approximately factorises into a product of one-dimensional distributions.}
{We find that the cosmic shear likelihood is significantly non-Gaussian. This leads to both a shift of the maximum of the posterior distribution and a significantly smaller credible region compared to the Gaussian case.  We re-analyse the CDFS cosmic shear data using the non-Gaussian likelihood in combination with conservative galaxy selection criteria that minimise calibration uncertainties. Assuming that the CDFS is a random pointing, we find $\sigma_8=0.68_{-0.16}^{+0.09}$ for fixed $\Omega_{\rm m}=0.25$. In a WMAP5-like cosmology, a value equal to or lower than this would be expected in $\approx 5\%$ of the times.
Taking biases into account arising from the way the CDFS was selected, which we model as being dependent on the number of haloes in the CDFS, we obtain $\sigma_8  = 0.71^{+0.10}_{-0.15}$. Combining the CDFS data with the parameter constraints from WMAP5 yields $\Omega_{\rm m}  = 0.26^{+0.03}_{-0.02}$ and $\sigma_8  = 0.79^{+0.04}_{-0.03}$ for a flat universe. 
}
{}

\maketitle

\section{Introduction}

Weak gravitational lensing by the large-scale structure in the Universe, or cosmic shear, is becoming a more and more important tool to constrain cosmological parameters. It is largely complementary to other cosmological probes like the cosmic microwave background or the clustering of galaxies, and particularly sensitive to the matter density $\Omega_{\rm m}$ and the normalisation of the matter power spectrum $\sigma_8$. Important constraints have already been  obtained by \citet{jonben}, who compiled a set of five weak lensing surveys, and from the CFHT Legacy Survey \citep{sembolonCFHT, semboloni06, fu08}. In subsequent years, a new generation of surveys like KIDS or  Pan-STARRS \citep{panstarrs} will allow cosmic shear to be measured with statistical uncertainties that are much smaller than the systematic errors both on the observational and the theoretical sides. Strong efforts are now being made to find sources of systematics in the process of shape measurement and shear estimation \citep[e.g.][]{mhb07}. In addition, new methods of shape measurement are being explored, such as the shapelet formalism \citep{refregier_shapelets,kuijken06} or the methods proposed in \citet{bernsteinjarvis} and \citet{miller07}.

It is equally important to have accurate theoretical model predictions that can be fit to the expected high-quality measurements. Currently, these models are all based on fitting formulae for the three-dimensional matter power spectrum derived from $N$-body simulations as given by \citet{pdps} and more recently by \citet{smithps}. However, these are only accurate at best to the percent level on the scales relevant to this and similar works when compared to ray-tracing simulations based on state-of-the-art $N$-body simulations \citep{mrrt}, such as the Millennium Run \citep{mrpaper}. Therefore, there is a strong need for a large ray-tracing effort to obtain accurate semi-numerical predictions for a range of cosmological parameters.

While a tremendous effort is currently being directed to the solution of these problems, the actual process of parameter estimation has so far received relatively little attention. Obviously, the statistical data analysis has to achieve the same accuracy as the data acquisition if the aforementioned efforts are not to be wasted.

The standard procedure for converting measurements of second-order cosmic shear statistics into constraints on cosmological parameters is to write down a likelihood function and to determine the location of its maximum for obtaining estimates of the cosmological parameters of interest. To make this feasible, several approximations are commonly made. Despite the shear field being non-Gaussian due to nonlinear structure growth, lacking an analytical description the likelihood is most often approximated by a multivariate Gaussian distribution. The covariance matrix for the Gaussian likelihood then remains to be determined, which is an intricate issue by itself.

In most previous studies, the dependence of the covariance matrix on cosmological parameters has been ignored when writing down the likelihood function. Instead, it was kept fixed to some fiducial cosmological model. The dependence of the covariance matrix on the cosmological parameters has been investigated in \citet{eifler08} for the case of Gaussian shear fields. The authors find that this has a significant effect on the constraints on cosmological parameters (reducing the size of the credible regions) and will be particularly important for future large-area surveys.

There are several approaches to determine the covariance for the fiducial set of parameters: \citet{sembolonCFHT} use the covariance matrix derived for a Gaussian shear field. Although this is rather easy to compute \citep{joachimicov}, the errors are strongly underestimated particularly on small scales. Another option is to estimate the covariance from the data itself \citep[e.g.][]{masseyCOSMOS}. This will become sensible and feasible mostly for the upcoming large surveys, which can be safely split into smaller subfields without severely underestimating cosmic variance. A third possibility, which currently seems to be the most accurate, is to measure the covariance matrix from a large sample of ray-tracing simulations. \citet{sembolon_cov} have provided a fitting formula which allows one to transform covariances computed for Gaussian shear fields into covariances including non-Gaussianity. Another promising way, which would also easily allow one to take into account the dependence on cosmological parameters, is the semi-analytical computation using the halo model \citep{scocc99,CoorayHuHaloCov,takada08}.

However, all these works are based on the assumption that the likelihood is well approximated by a Gaussian. In this paper, we study the impact of this assumption on the shape of the posterior probability distribution of the matter density parameter $\Omega_{\rm m}$ and the power spectrum normalisation $\sigma_8$. Furthermore, we compute Fisher matrix constraints for the four-dimensional parameter space spanned by $\Omega_{\rm m}$, $\sigma_8$, $h_{100}$ and $\Omega_\Lambda$. We propose a method to numerically compute the likelihood function from a large set of ray-tracing simulations based on the technique of \emph{independent component analysis} \citep[ICA, e.g.][]{ica1,ica2}. ICA is a technique for the separation of independent source signals underlying a set of observed random variables, a statistical method related to factor analysis and principal component analysis (PCA). An approach similar to ours, called \emph{projection pursuit density estimation}, which we use to verify our results, was proposed by \citet{PPDE}.

In their cosmic shear analysis of the combined HST GEMS and GOODS data of the \textit{Chandra} Deep Field South, \citet{schrabbaACS} (\citetalias{schrabbaACS} from hereon) have found a very low value of $\sigma_8(\Omega_{\rm m}=0.3) = 0.52 ^{+0.11}_{-0.15}$. In the second part of this paper, we present a re-analysis of the cosmic shear data of \citetalias{schrabbaACS}. Using our estimate of the non-Gaussian likelihood, we investigate whether cosmic variance alone is responsible for producing the low $\sigma_8$-estimate or whether the criteria applied by \citet{giacconi_cdfs} to select a field suitable for deep X-ray observations have a share in this.

The outline of our paper is as follows: in \secref{sec:ray-tracing}, we describe our sample of ray-tracing simulations which we use for the likelihood estimation. In \secref{sec:ICA}, we briefly review the lensing quantities relevant for this paper and Bayesian parameter estimation. We introduce our method of estimating the ``true'' likelihood and illustrate the impact of non-Gaussianity on parameter estimation using the example of a CDFS-like survey.
In Sec.\ \ref{sec:cdfs}, we present the improved cosmic shear analysis of the CDFS and investigate possible reasons for the low power spectrum normalisation found in \citetalias{schrabbaACS}.

\section{Ray-Tracing simulations \label{sec:ray-tracing}}
We have performed a set of 10 $N$-body simulations using the publically available code {\tt GADGET-2} \citep{GADGET-2}, all of which are realisations of the same WMAP-5-like cosmology ($\Omega_{\rm m}=0.25$, $\Omega_{\Lambda}=0.75$, $\Omega_{\rm b}=0.04$, $n_{\rm s}=1.0$, $\sigma_8=0.78$, $h_{100}=0.73$).
The simulation boxes are $L_{\rm box}=150\, h_{100}^{-1}{\rm Mpc}$ on a side, populated by $N_{\rm p}=256^3$ dark matter particles with masses of $m_{\rm p}=1.2\times 10^{10}\,h_{100}^{-1}\,M_\odot$.
We have started the simulations at \mbox{$z=50$} and obtained snapshots from $z=0$ to $z=4.5$ in intervals of $\Delta z$ corresponding to the box size, so that a suitable snapshot is available for each lens plane.

In the following, we only give a brief description of our ray-tracing algorithm and refer the reader to, for example,  \cite{jsw} or \citet{mrrt} for a more detailed introduction.

The ray-tracing is performed by dividing the dark matter distribution into redshift slices and projecting each slice onto a lens plane. Starting at the observer, light rays are shot through this array of lens planes. We assume that deflections only take place at the planes themselves, and that the rays propagate on straight lines in the space between two planes. In our case, each redshift slice corresponds to one output box of the $N$-body simulation and was projected as a whole onto a lens plane, preserving the periodic
boundary conditions of the simulation box. To avoid repetition of structure
along the line of sight, the planes were randomly shifted and rotated.
The light rays are shot from the observer through the set of lens planes,
forming a regular grid on the first plane. We then use FFT methods to compute
the lensing potential on each lens plane, from which we obtain the deflection
angle and its partial derivatives on a grid. The ray position and the
Jacobian of the lens mapping for each ray are obtained by recursion:
given the ray position on the current lens plane, its propagation direction
(known from the position on the previous plane), and the deflection angle on the
current plane interpolated onto the ray, we immediately obtain the ray
position on the next plane. Differentiation of this recursion formula with
respect to the image plane coordinates yields a similar relation for the
Jacobian of the lens mapping, which takes into account the previously computed tidal deflection field \citep[for a detailed description of the formalism used, see ][]{mrrt}. The recursion is performed until we reach the redshift cut-off at $z=4.5$.

We obtain the final Jacobian for a given source redshift distribution by performing a weighted average over the Jacobians for the light paths to each lens plane.
Since our aim is to create mock catalogues comparable to those of the CDFS field, we use the redshift distribution found for our revised galaxy catalogues \citep[][ see Sec.~\ref{sec:cdfs:data}]{smailpz}:
$$
p(z_{\rm s}) =	A\,\left(\frac{z_{\rm s}}{z_0}\right)^\alpha \exp\left[-\left(\frac{z_{\rm s}}{z_0}\right)^\beta\right]\;,
$$
where $z_0=1.55$, $\alpha=0.59$, $\beta=1.35$ and $A$ is a normalisation constant. This corresponds to a mean source redshift of $\bar z_{\rm s} = 1.54$.
We then create the mock source catalogue by randomly sampling the resulting shear maps with $N_{\rm s}=n_{\rm s}\Omega^2$ galaxies, where $n_{\rm s}=68\;{\rm arcmin}^{-2}$ is the number density of sources and $\Omega=0\fdg 5$ is the side length of the simulated field.
In total, we have produced $9600$ quasi-independent realisations of the CDFS field, based on different random shifts and rotations of the lens planes and the various $N$-body simulations.

\section{The non-Gaussianity of the cosmic shear likelihood \label{sec:ICA}}

\subsection{Cosmic shear \label{sec:cs}}
Perhaps the most common way to extract the lensing information from the measured shapes of distant galaxies is to estimate the two-point correlation functions of the distortion field. One defines two shear correlation functions \citep[for more details, see e.g.][]{saasfee3}
\begin{equation} \label{eq:shearcf_def}
	\xi_\pm(\theta) = \left< \epsilon_{\rm t}(\vec{\vartheta}) \epsilon_{\rm t}(\vec{\theta + \vartheta})\right> \pm \left< \epsilon_\times(\vec{\vartheta}) \epsilon_\times(\vec{\theta + \vartheta}) \right>\;\;,
\end{equation}
where $\epsilon_{{\rm t},\times}$ are the tangential and cross components of the measured ellipticity relative to the line connecting the two galaxies, and $\theta$ is the angular separation.
An unbiased estimator for the shear correlation functions for a random distribution of galaxies is given in \citet{schneidermap}:
\begin{equation} \label{eq:shearcf_estim}
	\hat\xi_\pm(\theta) = \frac{1}{N_{\rm p}(\theta)} \,\sum_{i,j}	\left( \epsilon_{i \rm t} \epsilon_{j\rm t} \pm \epsilon_{i \times} \epsilon_{j \times} \right)\;\Delta_\theta(|\vec\vartheta_i - \vec\vartheta_j|)\;\;.
\end{equation}
Here, $i$ and $j$ label galaxies at angular positions $\vec\vartheta_i$ and $\vec\vartheta_j$, respectively.
The function $\Delta_\theta(\phi)$ is $1$ if $\phi$ falls into the angular separation bin centred on $\theta$, and is zero otherwise. Finally, $N_{\rm p}$ is the number of pairs of galaxies in the bin under consideration.

\subsection{Parameter estimation}
Let us assume that we have measured the shear correlation functions $\xi_\pm(\theta_i)$ on $p/2$ angular separation bins ${\theta}_i$ and now wish to infer some parameters $\vec{\pi}$ of our model $\vec{m}(\vec{\pi})$ for $\xi_\pm({\theta}_i)$.
For what follows, we define the joint data vector $\vec{\xi} = ( \vec{\xi}_+, \; \vec{\xi}_-)^{\rm t}$, which in total is supposed to have $p$ entries.

Adopting a Bayesian point of view, our aim is to compute the posterior likelihood, i.e.\ the probability distribution of a parameter vector $\vec{\pi}$ given the information provided by the data $\vec{\xi}$:
\begin{equation}
	p(\vec{\pi} | \vec{\xi}) = \frac{p(\vec{\pi})}{p(\vec{\xi})} p(\vec{\xi} | \vec{\pi})\; .
\end{equation}
Here, $p(\vec{\pi})$ is the prior distribution of the parameters, which incorporates our knowledge about $\vec{\pi}$ prior to looking at the data; such can originate from previous measurements or theoretical arguments. The evidence $p(\vec{\xi})$ in this context simply serves as a normalisation factor. Hitherto, it has been assumed that the likelihood $p(\vec\xi | \vec{\pi})$ is a Gaussian distribution:
\begin{equation} \label{eq:gausslike} \begin{split}
	p(\vec{\xi} | \vec{\pi}) =& \frac{1}{(2\pi)^{p/2}\det \tens C(\vec \pi)^{1/2}} \\
		&\times\exp\left\{-\frac{1}{2} \left[\vec{\xi}-\vec{m}(\vec{\pi})\right]^{\rm t}\, \tens{C}^{-1}(\vec{\pi})\, \left[\vec{\xi}-\vec{m}(\vec{\pi})\right] \right\}\;\; ,
	 \end{split}
\end{equation}
where $\tens{C}(\vec{\pi})$ is the covariance matrix of $\vec{\xi}$ as predicted by the underlying model. Usually, however, the dependence of the covariance matrix upon cosmological parameters is not taken into account. Rather, the covariance that is computed for a fixed fiducial set of parameters $\vec\pi_0$ is used in Eq.~\eqref{eq:gausslike}. Under this approximation, the likelihood is a function of the difference $\vec\Delta(\vec{\pi})=\vec{\xi}-\vec{m}(\vec{\pi})$ only:
\begin{equation} \label{eq:fidulike}
	p(\vec{\xi} | \vec{\pi}) = L_{\vec{\pi}_0}\left[\vec{\Delta}(\vec{\pi})\right]\;\;.
\end{equation}

\subsection{Estimating the likelihood}\label{sec:likeest}
The choice of the functional form of the likelihood as given by Eq.~\eqref{eq:gausslike} is only approximate. Since the underlying shear field in the correlation function measurement becomes non-Gaussian in particular on small scales due to nonlinear structure formation, there is no good reason to expect the distribution of the shear correlation function to be Gaussian. Our aim therefore is to use a very large sample of ray-tracing simulations to estimate the likelihood and explore the effects of the deviations from a Gaussian shape on cosmological parameter constraints.

In this work, we have to sustain the approximation that the functional form of the likelihood does not depend on cosmology in order to keep computation time manageable.
Our ray-tracing simulations were all done for identical cosmological parameters, which is our fiducial parameter vector $\vec\pi_0$. Thus, as in Eq.~\eqref{eq:fidulike} the likelihood depends on cosmology only through the difference $\vec{\Delta}(\vec{\pi})=\vec{\xi}-\vec{m}(\vec{\pi})$.

Since $L_{\vec \pi_0}$ is the probability of obtaining the data $\vec{\xi}$ given the parameters $\vec{\pi}_0$, we in principle have to estimate the $p$-dimensional distribution of $\vec{\xi}$ from our sample of $N$ ray-tracing simulations. However, due to the high dimensionality of the problem, a brute force approach to estimate the full joint distribution is hopeless.
The problem would simplify considerably if we could find a transformation
\begin{equation}	\label{eq:sdef_general}
	\vec{s} = \vec{f}\left[\vec{\Delta}(\vec{\pi})\right]\;\; ,
\end{equation}
such that
\begin{equation} \label{eq:likefactor}
	p_s(\vec{s}|\vec{\pi}_0) = \prod_{i=1}^{n_{\rm IC}} \;p_{s_i}(s_i|\vec{\pi}_0)\;\; .
\end{equation}
Here, $\vec{f}$ is in general a mapping from $\mathbb{R}^p$ to $\mathbb{R}^{n_{\rm IC}}$ ($n_{\rm IC}\leq p$)  and $\vec{s}\in \mathbb{R}^{n_{\rm IC}}$ is our new data vector. This would reduce the problem to estimating $n_{\rm IC}$ one-dimensional probability distributions instead of a single $p$-dimensional one. Eq.\ \eqref{eq:likefactor} is equivalent to the statement that we are looking for a new set of basis vectors of $\mathbb{R}^{n_{\rm IC}}$ in which the components $s_i$ of the shear correlation function are statistically independent.
It is virtually impossible to find the (in general nonlinear) mapping $\vec{f}$. However, it is possible to make progress if we make the ansatz that $\vec{f}$ is linear:
\begin{equation}	\label{eq:sdef}
	\vec{s} = \tens{A}\vec{\Delta}(\vec{\pi})\;\; ,
\end{equation}
where $\tens{A}\in \mathbb{R}^{n_{\rm IC}\times p}$ is the transformation or ``un-mixing'' matrix.

Our likelihood estimation procedure is as follows: the first step is to remove first-order correlations from the data vector by performing a PCA \citep[e.g.~][]{nr}. This yields a basis in which the components of $\vec\xi$ are uncorrelated. If we knew that the distribution of $\vec\xi$ were Gaussian, this would be sufficient, because in this case uncorrelatedness is equivalent to statistical independence. However, for a general distribution, uncorrelatedness is only a necessary condition for independence. Since we suspect that the likelihood is non-Gaussian, a second change of basis, determined by the ICA technique (described in detail in the next section), is carried out which then results in the desired independence.
We then use a kernel density method \citep[see e.g.][ and references therein]{hastiebook,VenablesRipley} to estimate and tabulate the one-dimensional distributions $p_{s_i}(s_i|\vec{\pi}_0)$ in this new basis. The density estimate is constructed by smoothing the empirical distribution function of the observations of $s_i$,
\begin{equation}
	p^{\rm emp}_{s_i}(x) = \frac{1}{N} \sum_{j=1}^{N}\,\delta_{\rm D}({x-s_i^{(j)}})\;,
\end{equation}
where $s_i^{(j)}$ is the $j$-th of $N$ observations of $s_i$ and $\delta_{\rm D}$ is the Dirac delta-function, with a smooth kernel $K$. The estimate $\hat{p}_{s_i}$ of the desired density $p_{s_i}$ then is given by
\begin{equation} \label{eq:kerndensest}
	\hat{p}_{s_i}(x) = \frac{1}{Nb}\sum_{j=1}^{N}\, K\left( \frac{x-s_i^{(j)}}{b}\right)\;\;,
\end{equation}
where $s_i^{(j)}$ is the $j$-th of $N$ observations of $s_i$ and $b$ is the bandwidth. For the kernel $K$ we use a Gaussian distribution. It has been shown that the shape of the Kernel $K$ is of secondary importance for the quality of the density estimate; much more important is the choice of the bandwidth $b$. If $b$ is too small, $\hat{p}_{s_i}$ is essentially unbiased, but tends to have a high variance because the noise is not properly smoothed out. On the other hand, choosing a bandwidth that is too large results in a smooth estimate with low variance, but a higher bias, because real small scale features of the probability density are smeared out. Our choice of the bandwidth is based on the ``rule of thumb'' \citep[e.g.~][]{silverman86,scott92,davison03}: $b=0.9 \min(\hat\sigma,\,R/1.34)\,N^{-1/5}$. Here, $\hat\sigma$ is the sample standard deviation and $R$ is the inter-quartile range of the sample.

Constraints on cosmological parameters can now be derived as follows: we transform our set of model vectors and the measured correlation function to the new ICA basis:
\begin{equation}
	\breve{\vec{m}}(\vec{\pi}) = \tens{A}\,\vec{m}(\vec{\pi})\;\;,
\end{equation}
\begin{equation}
	\breve{\vec{\xi}} = \tens{A}\,\vec{\xi}\;\;,
\end{equation}
so that $\vec{s} = \breve{\vec{\xi}}-\breve{\vec{m}}(\vec{\pi})$.
The ICA posterior distribution is then given by
\begin{equation}
	p(\vec{\pi} | \vec{\xi}) \propto  p(\vec{\pi})\; \prod_{i=1}^{n_{\rm IC}} \;p_{s_i}(\breve{\xi}_i-\breve{m}_i(\vec{\pi})|\vec{\pi}_0)\;\;.
\end{equation}

\subsection{Independent Component Analysis}
We now briefly outline the ICA method \citep{ICAbook, ICApaper}, which we use to find the new basis in $\mathbb{R}^{n_{\rm IC}}$ in which the components of $\vec\Delta$ are (approximately) statistically independent.
ICA is best introduced by assuming that the data at hand were generated by the following linear model:
\begin{equation} \label{eq:ICAmodel}
	\vec{\Delta} = \tens{M}\vec{s}\;,
\end{equation}
where $\vec{s}$ is a vector of statistically independent source signals with non-Gaussian probability distributions and $\tens{M}$ is the mixing matrix. For simplicity, we will from now on only consider the case $n_{\rm IC}=p$, in which case the mixing matrix $\tens M$ is simply the inverse of the un-mixing matrix $\tens A$ in Eq.~\eqref{eq:sdef}.
The goal of ICA is to estimate both $\tens{M}$ and $\vec{s}$ from the data.

An intuitive, though slightly hand-waving way to understand how ICA works is to note that
a set of linear combinations $Y_i$ of independent, non-Gaussian random variables $X_j$ will usually have distributions that are more Gaussian than the original distributions of the $X_j$ (Central Limit Theorem). Reversing this argument, this suggests that the $X_j$ could be recovered from a sample of the $Y_i$ by looking for linear combinations of the $Y_i$ that have the least Gaussian distributions. These linear combinations will also be close to statistically independent.
A more rigorous justification of the method can be found in \citet{ICAbook}.

The ICA algorithm consists of two parts, the first of which is a preprocessing step: after subtracting the mean $\bar{\vec{\Delta}}=\langle \vec{\Delta} \rangle$ from $\vec{\Delta}$, the data is whitened, i.e.\ a linear transformation $\tilde{\vec{\Delta}}=\tens{L}(\vec{\Delta}-\bar{\vec{\Delta}})$ is introduced such that $\langle \tilde{\vec{\Delta}} \tilde{\vec{\Delta}}^{\rm t} \rangle=\tens{E}$, where $\tens{E}$ is the unit matrix. This can be achieved by the eigen-decomposition of the covariance matrix $\tens{C}=\tens{U}\tens{D}\tens{U}^{\rm t}$ of $\vec\Delta$, where $\tens{D}={\rm diag}(d_1,\ldots,d_p)$, by choosing $\tens{L}= \tens{D}^{-1/2}\tens{U}^{\rm t}$. Note that $\tens{U}$ is orthonormal and that $d_i\geq 0$ for all $i$. As will be discussed below, each source signal $s_i$ can only be determined up to a multiplicative constant using ICA. We choose these factors such that $\langle \vec{s}\vec{s}^{\rm t}\rangle = \tens{E}$. The effect of the whitening is that the new mixing matrix $\tilde{\tens{M}}=\tens{L}\tens{M}$ between $\tilde{\vec{\Delta}}$ and $\vec s$ is orthogonal. This can be seen as follows: $\tens{E} = \langle \tilde{\vec{\Delta}} \tilde{\vec{\Delta}}^{\rm t} \rangle = \tilde{\tens{M}} \langle \vec{s}\vec{s}^{\rm t} \rangle \tilde{\tens{M}}^{\rm t}$. Since we have chosen $\langle \vec{s}\vec{s}^{\rm t}\rangle = \tens{E}$, the claim follows.

After the preprocessing, the components of $\tilde{\vec{\Delta}}$ are uncorrelated. This would be equivalent to statistical independence if their distributions were Gaussian. However, as this is not the case here, a further step is needed. It consists of finding a new set of orthogonal vectors $\vec{w}_i$ (the row vectors of $\tilde{\tens{M}}$) such that the distributions $p_{z_i}(z_i)$ of
\begin{equation} \label{eq:ICAprojection}
	z_i=\tilde{\vec{\Delta}}_i\cdot\vec{w}_i
\end{equation}
maximise a suitable measure of non-Gaussianity. A common method to achieve this is to minimise the entropy (or approximations thereof) of the $z_i$, which is defined by
\begin{equation} \label{eq:entropy}
H_{z_i}=-\int \dd y \; p_{z_i}(y) \log p_{z_i}(y)\;.
\end{equation}
Since it can be shown that the Gaussian distribution has the largest entropy of all distributions of equal variance, this can be rewritten as maximising the so-called negentropy of the $z_i$, defined by
\begin{equation} \label{eq:negent}
	J_{z_i} = H_{z_i^{\rm Gauss}} - H_{z_i}\;.
\end{equation}
Here, $z_i^{\rm Gauss}$ is a Gaussian random variable with the same variance as $z_i$ and $J\left(z_i\right)\geq 0$.
Starting from randomly chosen initial directions $\vec{w}_i$, the algorithm tries to maximise $J\left(z_i\right)$ iteratively (in practice, it is sufficient to use a simple approximation to the negentropy). For more details, the reader is again referred to \citet{ICAbook}.

ICA suffers from several ambiguities, none of which, however, is crucial for this work. First of all, the amplitudes of the source signals cannot be determined, since any prefactor $\lambda$ to the signal $s_i$ can be cancelled by multiplication of the corresponding column of the mixing matrix by $1/\lambda$. Secondly, the order of the independent components is not determined, since any permutation of the $s_i$ can be accommodated by corresponding changes to $\tens{M}$. Thirdly, ICA does not yield a unique answer if at least some of the $s_i$ are Gaussian -- the subset of Gaussian signals is only determined up to an orthogonal transformation.
This is not an issue in our context, since the Gaussian signals will be uncorrelated thanks to the preprocessing steps, and uncorrelatedness implies statistical independence for Gaussian random variables.

Several interpretations of ICA and algorithms exist and are described in detail in \citet{ICAbook}.
In this work, we use an implementation of the {\tt fastICA} algorithm \citep{fastica} for the {\tt R} language \citep{Rcit}\footnote{{\tt http://www.r-project.org/}}.

\subsection{Tests}
In this section, we present the results of a number of tests we have performed to insure that our results are not affected by convergence issues or statistical biases of any kind.\\

The {\tt fastICA} algorithm requires a set of randomly chosen directions $\vec{w}_i$ as initial conditions. It then iteratively computes corrections to these vectors in order to increase the negentropy of the projections of the data vectors onto these directions (Eq.~\ref{eq:ICAprojection}), followed by an orthonormalisation step. It is not clear a priori whether the algorithm will settle in the same negentropy maxima for different sets of initial vectors. This concern is backed by the fact that at least some of the $p_{s_i}(s_i|\vec{\pi}_0)$ might be very close to Gaussian, which might hamper convergence even further.
We have therefore tested whether we obtain the same set of basis vectors from a large number of different initial conditions. We find that this is indeed the case for those $\vec{w}_i$ for which the distribution of $p_{z_i}(z_i=\tilde{\vec{\Delta}}\cdot\vec{w}_i)$ departs significantly from a Gaussian. As expected, the directions leading to a rather Gaussian $p_{z_i}$ are different for different starting values, reflecting the inability of ICA to distinguish between Gaussian source signals.
However, the posterior distributions derived using our algorithm do not differ notably when using different initial conditions. This is even true if the {\tt fastICA} algorithm does not formally converge (i.e. when the differences of some of the basis vectors between two iterations is not small): after a few hundred iterations, the non-Gaussian directions are determined and do not change anymore. The reason for not reaching convergence is that the algorithm still tries to find negentropy maxima in the subspace of Gaussian directions.

As has been noted in \citet{HartlapCov}, statistical biases can become significant already for the Gaussian approximation of the likelihood (Eq.~\ref{eq:gausslike}): care has to be taken if the covariance matrix of the correlation function (given on $p$ bins) is estimated from a finite set of $N$ simulations or observations. Inverting the estimated covariance yields a biased estimate of the inverse:
\begin{equation} \label{eq:covbias}
	\left\langle \hat{\tens{C}}^{-1}\right\rangle = \frac{N-1}{N-p-2}\; \tens{\Sigma}^{-1}\; \mbox{for}\; p<N-1\, ,
\end{equation}
where  $\hat{\tens{C}}$ is the estimated and $\tens{\Sigma}$ the true covariance matrix.
This bias leads to an underestimation of the size of credible regions by a factor of $(N-p-2)/(N-1)\approx 1-p/N$.
We suspect that a similar bias occurs in our likelihood estimation procedure.
\begin{figure}
\resizebox{\hsize}{!}{\includegraphics[angle=270]{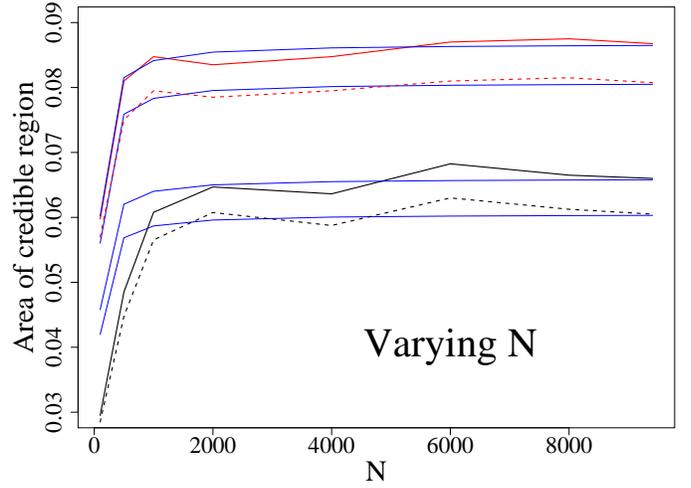}}
\caption{Area of the $68\%$ (dashed lines) and $99\%$ (solid lines) credible regions in the $\Omega_{\rm m}$-$\sigma_8$-plane as function of the sample size $N$, for the Gaussian likelihood (red, upper curves) and the likelihood computed using the ICA algorithm (black, lower curves). Blue lines are the predicted areas based on Eq.\ \eqref{eq:covbias}.}\label{fig:biasN}
\end{figure}
In Fig.~\ref{fig:biasN}, we therefore plot the area of the $68\%$ and $99\%$ credible regions of the posterior distribution for $\Omega_{\rm m}$ and $\sigma_8$ (keeping all other cosmological parameters fixed to their fiducial values) as functions of the number $N$ of observations of the correlation functions used to estimate the ICA transformation (black curves). To exclude noise effects from the analysis, we use the theoretical prediction of the correlation function for the fiducial cosmological parameters as data vector.
 We set $p=30$ throughout. For comparison, we also show the areas computed using the Gaussian likelihood (red curves). In the latter case, the bias predicted by Eq.~\eqref{eq:covbias} is clearly visible as a decrease of the area when $N$ becomes small. The ICA method  suffers from a similar bias, although the behaviour at small $N$ seems to be slightly different. More important, though, is the fact that this bias is unimportant for reasonably large sample sizes ($N \gtrsim 2000$). Since we always use the full sample ($N=9600$) in the following, this bias is completely negligible.

\begin{figure}
\resizebox{\hsize}{!}{\includegraphics[angle=270]{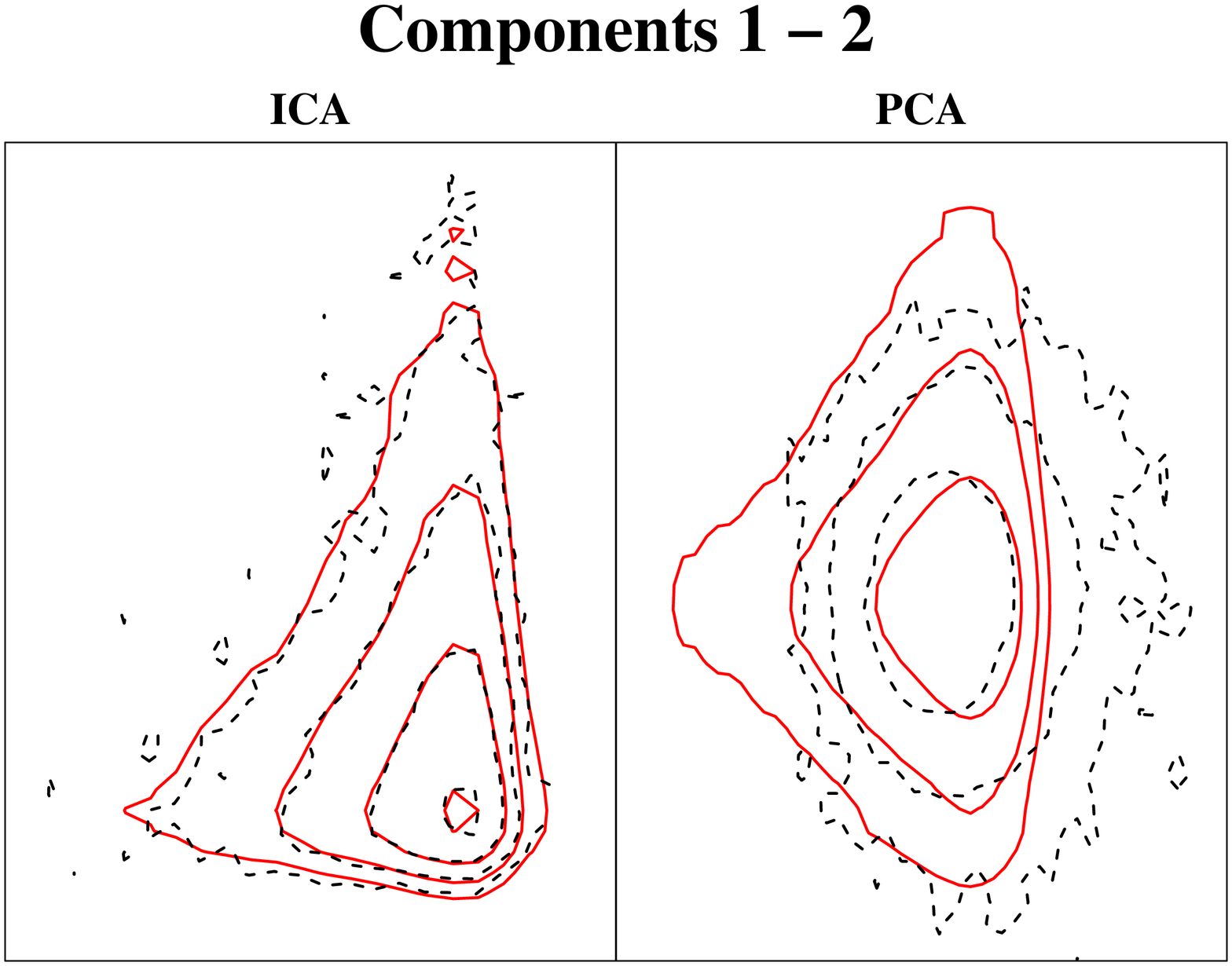}}

\vspace{0.5cm}

\resizebox{\hsize}{!}{\includegraphics[angle=270]{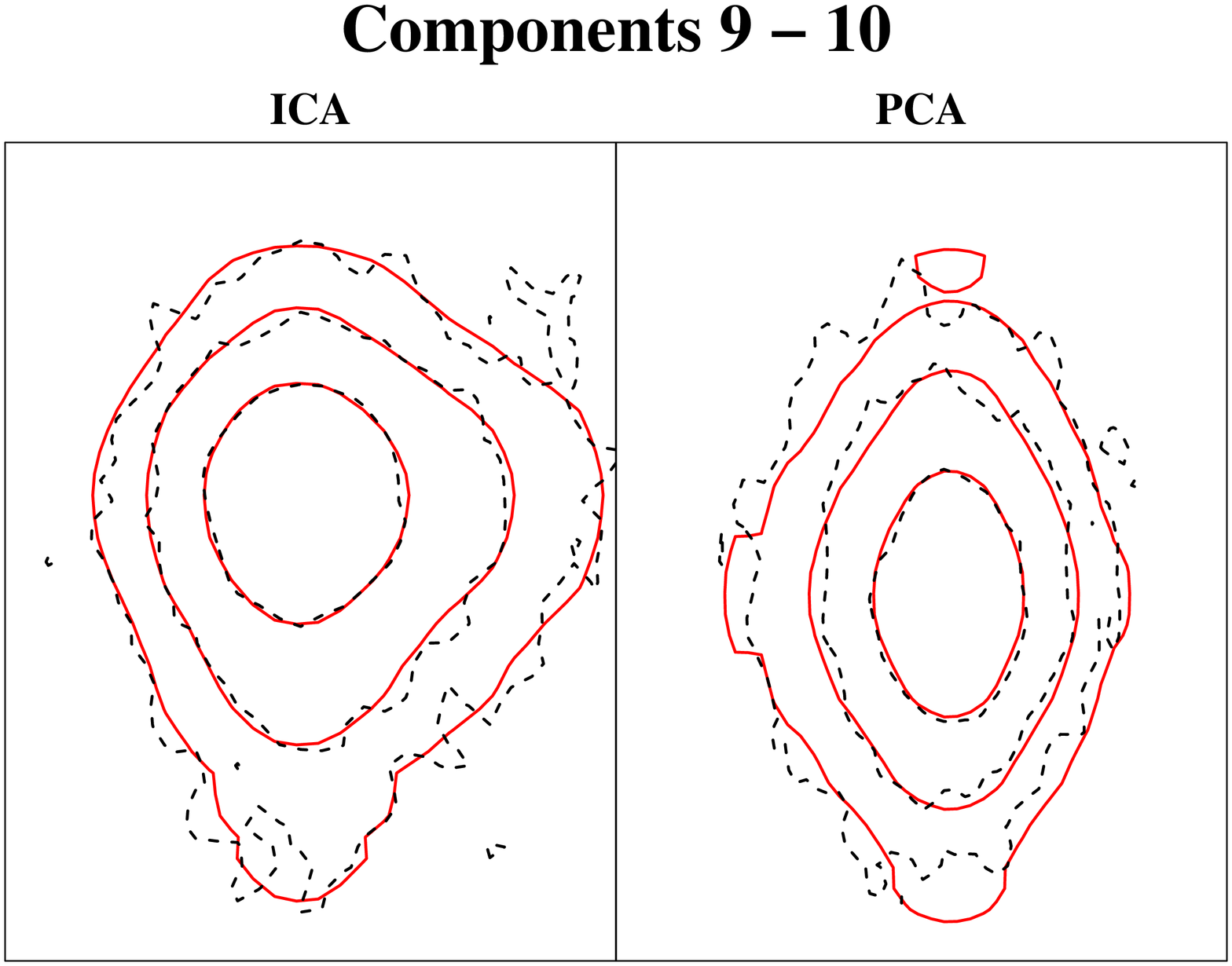}}
\caption{Comparison of the joint distributions $p(s_i,s_j)$ (black dashed contours) and the product $p_{s_i}(s_i)\,p_{s_j}(s_j)$ (solid red contours) for the two most non-Gaussian components ($i=1$, $j=2$) and two rather Gaussian ones ($i=9$, $j=10$), after performing ICA (left panels) and PCA (right panels). The components have been ranked and labelled according to their non-Gaussianity; the $i$-th PCA component is in general not the same as $i$-th ICA-component. In the right panel of each plot, the distributions with respect to the PCA basis vectors are shown and in the left panel, the distributions in the ICA basis are displayed. Statistical independence is indicated by $p(s_i,s_j)=p_{s_i}(s_i)\,p_{s_j}(s_j)$.}
\label{fig:pdf_indep}
\end{figure}
\begin{figure}
\resizebox{\hsize}{!}{\includegraphics[angle=270]{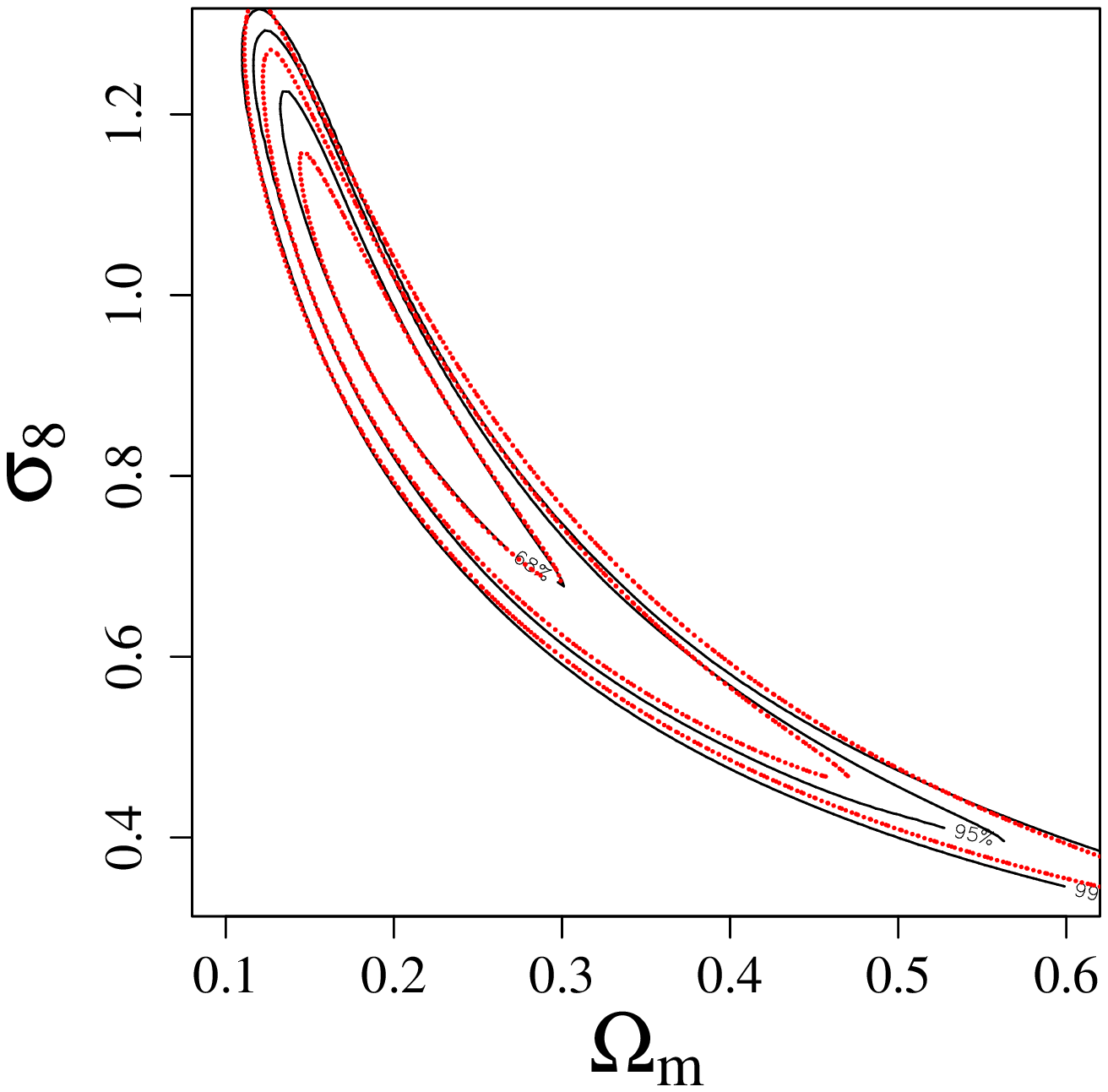}}
\caption{Comparison of the posterior likelihoods for $(\Omega_{\rm m}, \sigma_8)$, computed using the ICA likelihood (black contours) and the PPDE likelihood (red contours). Shown are the contours of the $68\%$, $95\%$ and $99\%$ credible regions.}
\label{fig:ica_ppde}
\end{figure}

Our method to estimate the likelihood crucially depends on the assumption that a linear transformation makes the components of the shear correlation vectors statistically independent. A necessary condition for mutual statistical independence of all $s_i$ is pairwise independence. The components $i$ and $j$ are called pairwise statistical independent if $ p(s_i,s_j)=p_{s_i}(s_i)\,p_{s_j}(s_j)$. We therefore compare the joint pairwise distributions $p(s_i,s_j)$ to the product distributions $p_{s_i}(s_i)\,p_{s_j}(s_j)$, where we estimate $p(s_i,s_j)$ using a two-dimensional extension (using a bi-variate Gaussian kernel) of the kernel density method given by Eq.~\eqref{eq:kerndensest}. We give two examples in Fig.~\ref{fig:pdf_indep}, where we compare the joint and product distributions of the two most-non-Gaussian components and two nearly Gaussian components. As expected, a simple PCA is not enough to achieve pairwise statistical independence in the non-Gaussian case. Only after performing the ICA, pairwise independence is achieved.

A more rigorous test for mutual statistical independence for the multivariate, continuous case was proposed by \citet{ICA03:Chiu}.
It is based on the observation that if $x$ is a continuous random variable and $P(x)$ is its cumulative distribution function (CDF), then $z=P(x)$ is uniformly distributed in $[0,1]$. If we are given a set of statistically independent random variables $s_i$, this means that the joint distribution of $z_i=P_i(s_i)$, where again $P_i$ is the CDF of $s_i$, is uniform in the multidimensional unit cube. On the other hand,
if the assumption of statistical independence of the $s_i$ is violated, the joint density $p_{\vec z}$ of the $z_i$ is given by
\begin{eqnarray}
	p_{\vec z}(\vec z) &=& p_{\vec z}\left[ P_1(s_1),\ldots,P_n(s_n)\right]\nonumber\\
	&=& p_{\vec s} (s_1,\ldots,s_n)\,\left\lvert \frac{\partial\vec z}{\partial\vec s} \right\rvert^{-1}\nonumber\\
	&=& \frac{p_{\vec s} (s_1,\ldots,s_n)}{\prod_{i=1}^n\,p_i(s_i)}\;.
\end{eqnarray}
Here, $p_i(s_i)$ is the distribution function of $s_i$ only and $p_{\vec s}$ is the joint distribution function of $s_1,\ldots,s_n$.
This means that the joint distribution of the $z_i$ is not uniform if the $s_i$ are statistically dependent. Therefore, we can test if the $s_i$ we obtain from the ICA procedure are indeed independent by computing their empirical cumulative distribution functions, carrying out the above transformation and finally testing for multivariate uniformity. Such a test was described in \citet{uniftest}, to which we refer the reader for more details.
\begin{figure*}
\sidecaption
\includegraphics[width=12cm]{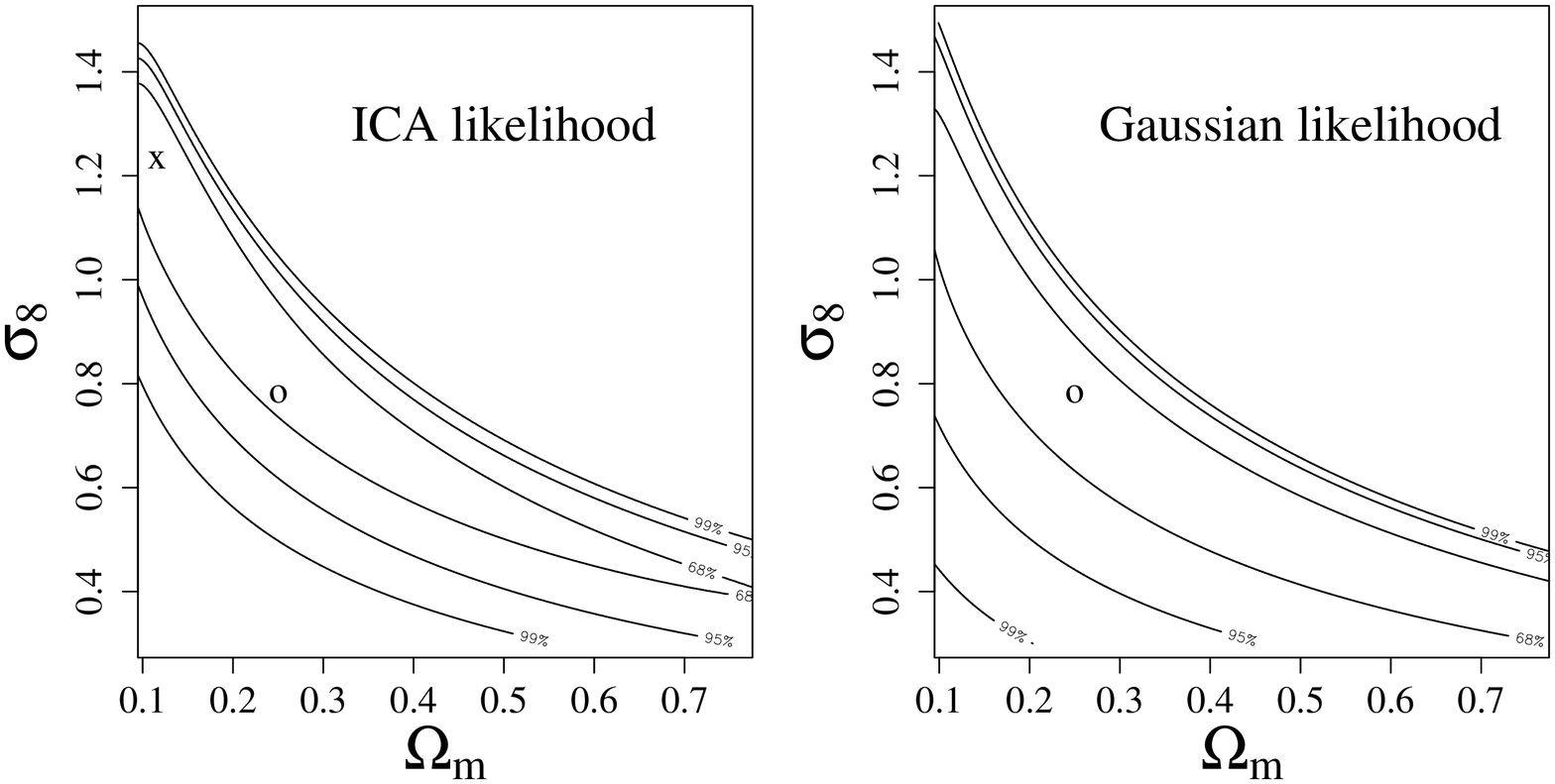}
\caption{Comparison of the posterior likelihoods for $(\Omega_{\rm m}, \sigma_8)$, computed using the ICA likelihood (left panel) and the Gaussian approximation (right panel). Shown are the contours of the $68\%$, $95\%$ and $99\%$ credible regions. The maximum of the ICA posterior is denoted by `x'; the maximum of the posterior based on the Gaussian likelihood coincides with the fiducial parameter set and is marked by the symbol `o'.}
\label{fig:2dcontours}
\end{figure*}

Applying the test to the $s_i$ that we have obtained from our ICA procedure, we have to reject statistical independence at $99\%$ confidence. This means that the ICA does not remove all dependencies between the components of the shear correlation function. This result, however, does not give an indication of how these residual dependencies affect our likelihood estimate and the conclusions regarding constraints on cosmological parameters.
We therefore compare the constraints derived from the ICA likelihood with the constraints from the likelihood estimated using an alternative method, called \emph{projection pursuit density estimation} \citep[PPDE;][]{PPDE}, which we describe in detail in App.~\ref{sec:ppde}. This method is free from any assumptions regarding statistical independence and therefore provides an an ideal cross-check for the ICA method.
For the comparison, we have computed the shear correlation functions with $p=10$, and we also use $n_{\rm IC}=10$ independent components. The resulting contours in the $\Omega_{\rm m}$-$\sigma_8$-plane are shown in Fig.~\ref{fig:ica_ppde}. Both posterior likelihoods are very similar, although the credible regions of the PPDE posterior have a slightly smaller area than the contours of the ICA posterior (which actually supports the findings presented in the next section). Given the good agreement of the two methods, we will henceforth only make use of the ICA procedure, which is considerably faster and numerically less contrived than PPDE.

\subsection{Results on the posterior}



\begin{figure*}
\sidecaption
\includegraphics[width=12cm]{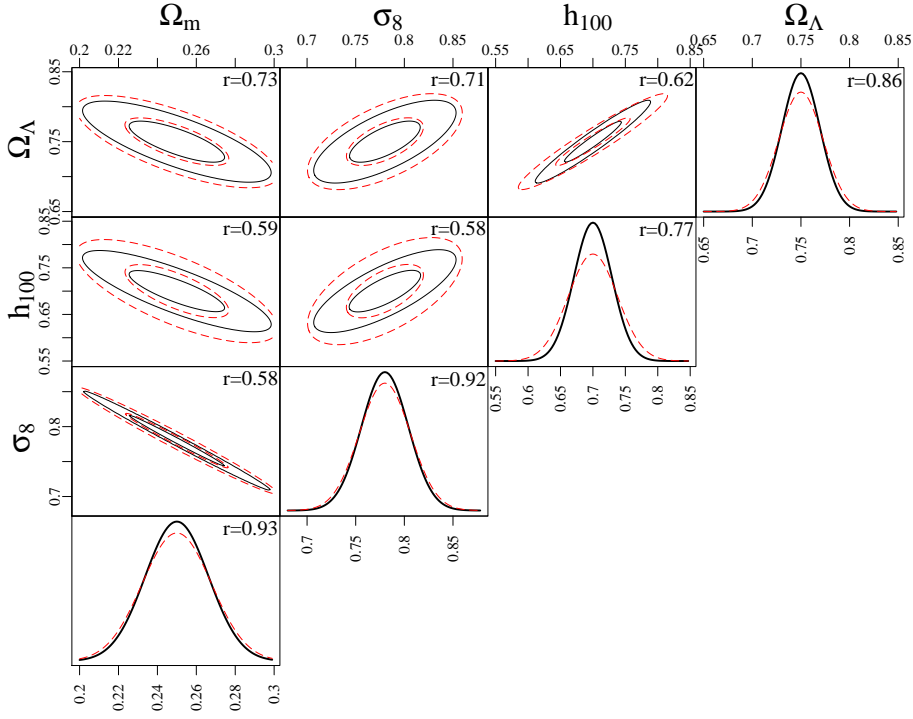}
\caption{Fisher matrix constraints for a hypothetical $1500$-${\rm deg}^2$ survey, consisting of $6000$ CDFS-like fields. The plots on the diagonal show the 1D marginals, the off-diagonal plots the 2D marginals derived from the full 4D posterior. The red dashed (black solid) lines/contours have been computed using the Fisher matrix of the Gaussian likelihood (the ICA likelihood).}
\label{fig:contours2}
\end{figure*}

The most interesting question is how much the posterior distribution computed from the non-Gaussian ICA likelihood will differ from the Gaussian approximation. We have investigated this for the case of the CDFS and the parameter set $(\Omega_{\rm m},\,\sigma_8)$.
Here and henceforth, we use 15 angular bins for $\xi_+$ and $\xi_-$ in the range from $12\arcsec$ to $30\arcmin$, i.e.~$p=30$. For the data vector, we do not use the correlation functions from our simulations, but take the theoretical prediction for our fiducial parameter set instead. This allows us to study the shape of the posterior likelihood independent of noise in the data and biases due to the fact that the theoretical model does not quite match the mean correlation function from the simulations. In Fig.\ \ref{fig:2dcontours}, we show the contours of the posterior computed in this way from the likelihood estimated using our ICA method (left panel) and from the Gaussian likelihood. We have assumed $\sigma_{|\epsilon|}=0.45$ for the dispersion of the intrinsic galaxy ellipticities.
The shape of the ICA posterior is different from that of the Gaussian approximation in three respects: it is steeper (leading to smaller credible regions), the maximum is shifted towards higher $\sigma_8$ and lower $\Omega_{\rm m}$, and the contours are slightly tilted. The first two differences can be traced back to the shape of the distributions of the individual ICA components: most of the distribution functions $p_{s_i}$ are generally slightly steeper than a Gaussian and most of the non-Gaussian components are in addition strongly skewed, thus shifting the peak of the posterior. Generally, these differences are more pronounced in the direction of the $\Omega_{\rm m}$--$\sigma_8$-degeneracy and towards lower values of both parameters, where the posterior is shallower.

Of more practical relevance is how the parameter constraints change when the ICA likelihood is used for the analysis of large weak lensing surveys. Here, we consider surveys consisting of $N_{\rm f}$ CDFS-like fields.
Bayesian theory states that if $N_{\rm f}$ is large enough, the posterior probability distribution of the parameters becomes Gaussian, centred on the true parameter values, with covariance matrix $(N_{\rm f}\,\tens{F})^{-1}$ \citep[e.g.~][]{bayesian_data_ananlysis_book}. Here, $\tens{F}$ is the Fisher matrix \citep{kendall_stuart}, which is defined by
\begin{equation} \label{eq:fisherdef}
\tens{F}_{\alpha\beta} = \left\langle \frac{\partial \log L}{\partial \pi_\alpha} \frac{\partial \log L}{\partial \pi_\beta} \right\rangle \;\; ,
\end{equation}
where $\langle\cdot\rangle$ denotes the expectation value with respect to the likelihood function.
If the likelihood is Gaussian and if the covariance matrix $\tens C$ does not depend on cosmology, one can show that
\begin{equation} \label{eq:gaussfisher}
	\tens{F}_{\alpha\beta} = \sum_{i,j} \tens{C}^{-1}_{ij}\,\frac{\partial m_i(\vec{\pi})}{\partial \pi_\alpha}\, \frac{\partial m_j(\vec{\pi})}{\partial \pi_\alpha}\;\;.
\end{equation}

Eq.~\eqref{eq:fisherdef} provides us with a way to estimate the Fisher matrix for the non-Gaussian likelihood. For each ray-tracing realisation of the CDFS, we compute the logarithm of the posterior distribution $\log p(\vec{\pi}|\vec{\xi})$ and its derivatives with respect to the cosmological parameters at the fiducial parameter values. Since we use uniform priors for all cosmological parameters, the derivatives of the log-posterior are identical to those of the log-likelihood. We can then compute the Fisher matrix by averaging over all realisations:
\begin{equation} \label{eq:icafisher_exp}
	\hat{\tens{F}}_{\alpha\beta} = \frac{1}{N}\,\sum_{k=1}^{N}\, \frac{\partial \log p(\vec{\pi}|\vec{\xi})}{\partial \pi_\alpha} \frac{\partial \log p(\vec{\pi}|\vec{\xi})}{\partial \pi_\beta}\;\;.
\end{equation}
In App.~\ref{sec:icafisher}, we show that the expression for the Fisher matrix of the ICA likelihood can be evaluated further to be
\begin{equation} \label{eq:icafisher_deriv}
	\tens{F}_{\alpha\beta} = \sum_{i}\; \frac{\partial \breve{m}_i}{\partial \pi_\alpha} \frac{\partial \breve{m}_i}{\partial \pi_\beta}\; \int\dd s_i\; p_{s_i}(s_i)\,\left(\frac{\partial \log p_{s_i}(s_i)}{\partial s_i}\right)^2 \;\;.
\end{equation}
This equation allows a simpler, alternative computation of $\tens{F}$ from the estimated $p_{s_i}(s_i)$, as discussed in App.~\ref{sec:icafisher}.

We have used Eqns.~\eqref{eq:gaussfisher} and \eqref{eq:icafisher_deriv} to compute the Fisher matrices for a $1500$-${\rm deg}^2$ survey ($N_{\rm f}= 6000$). We fit for four cosmological parameters ($\Omega_{\rm m},\,\sigma_8,h_{100},\,\Omega_\Lambda$), keeping all other parameters fixed to their true values.
To visualise the posterior, we compute two-dimensional marginalised posterior distributions for each parameter pair as well as the one-dimensional marginals for each parameter.
The results are shown in Fig.~\ref{fig:contours2}. A general feature of the ICA likelihood, which has already been apparent in the 2D-analysis (Fig.\ \ref{fig:2dcontours}), is that the credible intervals are significantly smaller than the ones derived from the Gaussian likelihood. For the two-dimensional marginal distributions, the area of the $68\%$ credible regions derived from the ICA likelihood are smaller by $\approx 30 - 40\%$. The one-dimensional constraints are tighter by $\approx 10 - 25\%$.
In addition we find that the ICA Fisher ellipses in some cases are slightly tilted with respect to those computed using the Gaussian likelihood. This is particularly apparent for parameter combinations involving the Hubble parameter. Note that the shift of the maximum observed in the two-dimensional case for a single CDFS-like field is absent here because it was assumed for the Fisher analysis that the posterior is centred on the true parameter values.

\section{How odd is the Chandra Deep Field South?} \label{sec:cdfs}

\subsection{The CDFS cosmic shear data} \label{sec:cdfs:data}
The second part of this work is based on the cosmological weak lensing analysis of the
combined HST GEMS and GOODS data of the CDFS \citep{rix04,Giavalisco04}, which was presented in \citetalias{schrabbaACS}. The mosaic comprises 78 ACS/WFC tiles imaged in F606W, covering a total area of \mbox{$\sim 28^\prime \times 28^\prime$}. We refer the reader to the original publication for details on the
data and weak lensing analysis, which applies the KSB+ formalism \citep{ksb95,luk97,hfk98}.

In \citetalias{schrabbaACS}, the cosmic shear analysis was performed using two different signal-to-noise and magnitude cuts. The first one selects galaxies with \mbox{$\mathrm{S}/\mathrm{N}>4$} and has no magnitude cut, and the second one applies a more conservative selection with \mbox{$\mathrm{S}/\mathrm{N}>5$} and \mbox{$m_{606}<27.0$}, where \mbox{$\mathrm{S}/\mathrm{N}$} is the shear measurement signal-to-noise ratio as defined in \citet{ewb01}.
The drizzling process in the data reduction introduces correlated noise in adjacent pixels. While these correlations are ignored in the computation of \mbox{$\mathrm{S}/\mathrm{N}$}, an approximate correction factor (see \citetalias{schrabbaACS}) is taken into account for \mbox{$\mathrm{S}/\mathrm{N}^\mathrm{true}$}, making the above cuts \mbox{$\mathrm{S}/\mathrm{N}^\mathrm{true}\gtrsim 1.9$} and \mbox{$\mathrm{S}/\mathrm{N}^\mathrm{true}\gtrsim 2.4$} respectively.
The two selection criteria yielded moderately different $\sigma_8$-estimates of $0.52^{+0.11}_{-0.15}$ and $0.59^{+0.11}_{-0.14}$ for \mbox{$\Omega_\mathrm{m}=0.3$} (median of the posterior), not assuming a flat Universe. The errors include the statistical and redshift uncertainties. This translates to $\sigma_8=0.57^{+0.12}_{-0.16}$ and $0.65^{+0.12}_{-0.15}$ for our fiducial cosmology with \mbox{$\Omega_\mathrm{m}=0.25$}.
The difference of the two estimates  was considered as a measure for the robustness and hence systematic accuracy of our shear measurement pipeline.
While the analysis of the ``Shear TEsting Programme 2'' (STEP2) image simulations \citep{mhb07} indicated no significant average shear calibration bias for our method, a detected dependence on galaxy magnitude and size could effectively bias a cosmic shear analysis through the redshift dependence of the shear signal \citep[see also][]{semboloni08}.
In order to better understand the difference between the two estimates found in \citetalias{schrabbaACS}, and to
exclude any remaining calibration uncertainty in the current analysis, we further investigate  the shear recovery accuracy as a function of the signal-to-noise ratio using the STEP2 simulations in Appendix \ref{se:ap:step2}.
Here we conclude that our KSB+ implementation under-estimates gravitational shear for very noisy galaxies with \mbox{$\mathrm{S}/\mathrm{N}^\mathrm{true}\lesssim 2.5$}, which likely explains the lower signal found in \citetalias{schrabbaACS} when all galaxies with \mbox{$\mathrm{S}/\mathrm{N}>4$} (\mbox{$\mathrm{S}/\mathrm{N}^\mathrm{true}\gtrsim 1.9$}) were considered.
For the more conservative selection criteria we find no significant mean shear calibration bias and a variation as a function of magnitude and size of  $\lesssim\pm5\%$. Therefore we base our current analysis on the more robust galaxy sample with  \mbox{$\mathrm{S}/\mathrm{N}>5$} (\mbox{$\mathrm{S}/\mathrm{N}^\mathrm{true}\gtrsim 2.4$}) and \mbox{$m_{606}<27.0$}, which yields a galaxy number density of $68\,\mathrm{arcmin}^{-2}$. Based on the simulations, any remaining calibration uncertainty should be negligible compared to the statistical uncertainty.

Note that \citet{heymans05} found a higher estimate of $\sigma_8(\Omega_{\rm m}/0.3)^{0.65}=0.68\pm 0.13$ from GEMS, where they extrapolated the redshift distribution from the relatively shallow COMBO-17 photometric redshifts \citep{wolf04}. Using deeper data from the GOODS-MUSIC sample \citep{grazian06}, \citetalias{schrabbaACS} were able to show that the COMBO-17 extrapolation significantly underestimates the mean redshift for GEMS, leading to the difference in the results for $\sigma_8$.

\begin{figure}
\resizebox{\hsize}{!}{\includegraphics[]{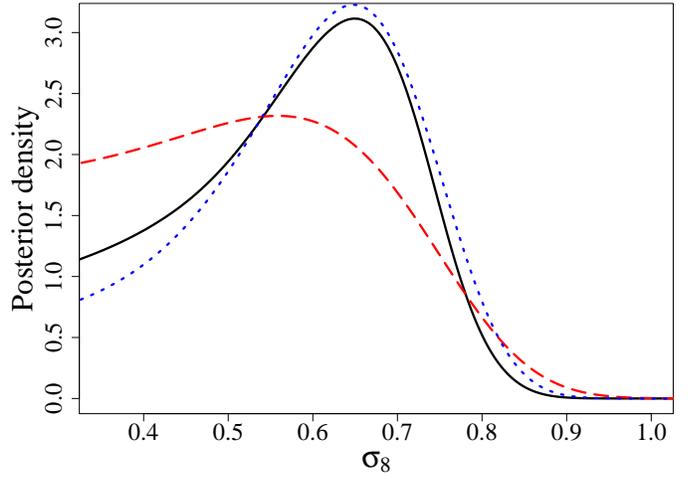}}
\caption{Posterior distributions for $\sigma_8$ as computed from the CDFS data. The black solid line corresponds to the ICA likelihood, the red dashed line is from the Gaussian likelihood whose covariance matrix was estimated from the ray-tracing simulations. The blue dotted line was computed from the Gaussian likelihood with an analytically computed covariance matrix, assuming that the shear field is Gaussian. The similarity of the posterior densities derived from the ICA likelihood and using the Gaussian covariance matrix is purely coincidental, occurring only for this particular data vector.}
\label{fig:cdfs_s8_fit}
\end{figure}

In Fig.~\ref{fig:cdfs_s8_fit}, we show the posterior distribution for $\sigma_8$ based on this sample of galaxies. For the fit, all other cosmological parameters were held fixed at the fiducial values chosen for our ray-tracing simulations. This avoids complications in the discussion of cosmic variance and field selection biases due to the effect of parameter degeneracies. We choose a flat prior for $\sigma_8$, with a lower boundary of $\sigma_{8,\,{\rm min}}=0.35$ to cut off the tail of the posterior distribution towards small values of the power spectrum normalisation, which is caused by the fact that the difference (and therefore the likelihood) between the data and the model vectors changes only very little when $\sigma_8$ (and therefore the shear correlation function) is very small. We have performed the fit for the ICA likelihood as well as for the Gaussian approximation to the likelihood. For the latter, the covariance matrix was in one case estimated from the full sample of our ray-tracing simulations, and in the other case computed analytically assuming that the shear field is a Gaussian random field \citep{joachimicov}.
The striking similarity of the posterior densities derived from the ICA likelihood and using the Gaussian covariance matrix for this particular data vector is merely a coincidence and is in general not seen for our set of simulated correlation functions.\\
For estimates of $\sigma_8$, we use the maximum of the posterior (henceforth we write ICA-MAP for the maximum of the non-Gaussian likelihood, and Gauss-MAP if the Gaussian approximation is used), although we also quote the median (ICA median) for comparison with \citetalias{schrabbaACS}. In the first case, our credible intervals are highest posterior density intervals, whereas for the median we choose to report the interval for which the probability of $\sigma_8$ of being below the lower interval boundary is as high as being above the upper boundary. The results are summarised in Tab.~\ref{tab:cdfs_results}.

\begin{table}
	\caption{Estimates of $\sigma_8$ from the CDFS}
	\label{tab:cdfs_results}
	\centering
	\begin{tabular}{c c c c}
		\hline\hline
		&ICA likel. & Gaussian likel. &  Gaussian likel.\\
		& & (ray-tracing cov.)&(Gaussian cov.)\\
		\hline
		MAP & $0.68_{-0.16}^{+0.09}$ & $0.59_{-0.19}^{+0.10}$ & $0.68_{-0.14}^{+0.10}$\\
		Median & $0.62_{-0.11}^{+0.11}$ & $0.57_{-0.15}^{+0.15}$ & $0.64_{-0.14}^{+0.10}$\\
		\hline
	\end{tabular}
\end{table}

\subsection{Cosmic Variance}
The original estimates for $\sigma_8$ given in \citetalias{schrabbaACS} and those found in the previous section for the Gaussian likelihood are rather low compared to the value reported by WMAP5 \citep{dunkley09}. This problem appears less severe when the full non-Gaussian likelihood is used, but the $\sigma_8$-estimate is still rather low.
It is therefore interesting to know whether this can be fully attributed to cosmic variance or whether the way in which the CDFS was originally selected biases our estimates low.

To begin, we determine the probability of finding a low $\sigma_8$ in a CDFS-like field if the pointing is completely random. We estimate the sampling distribution of the $\sigma_8$-MAP estimators for Gaussian and ICA likelihoods from the full sample of our ray-tracing simulations.
We compute the posterior likelihood for $\sigma_8$ using a uniform prior in the range $\sigma_8 \in [0.35;\,1.8]$ and determine the MAP estimator $\hat{\sigma}_8$. As in the previous sections, we do this using both the Gaussian and the ICA likelihoods. To separate possible biases of the estimators from biases that might arise because the model prediction based on \citet{smithps} does not quite fit our simulations, we correct the simulated correlation functions for this: if $\vec{\xi}^{(i)}$ is the correlation function measured in the $i$-th realisation, then
\begin{equation} \label{eq:recenter}
	\vec{\xi}_{\rm rc}^{(i)} = \vec{\xi}^{(i)} - \left\langle \vec{\xi} \right\rangle + \vec{m}(\vec{\pi}_0)\;,
\end{equation}
is the ``re-centred'' shear correlation, where $\left\langle \vec{\xi} \right\rangle$ is the mean of all realisations and $\vec{m}(\vec{\pi}_0)$ is our fiducial model.

\begin{figure}
\resizebox{\hsize}{!}{\includegraphics[]{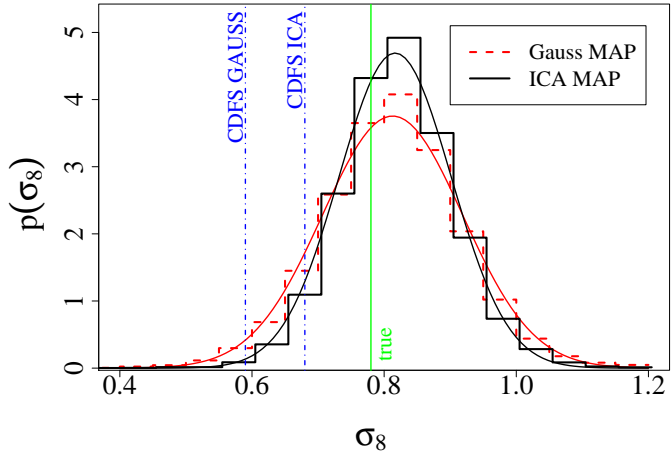}}
\caption{Sampling distributions of the MAP estimators of $\sigma_8$, derived from 9600 realisations of the CDFS. All other parameters were held fixed at their fiducial values for the fit. The histogram with red dashed lines has been obtained from the Gaussian likelihood, the one with solid lines from the ICA likelihood. Also shown are the best fitting Gaussian distributions. We indicate the fiducial value of $\sigma_8$ and our estimates from the CDFS with vertical lines.}
\label{fig:cdfs_s8}
\end{figure}

\begin{figure}
\resizebox{\hsize}{!}{\includegraphics[]{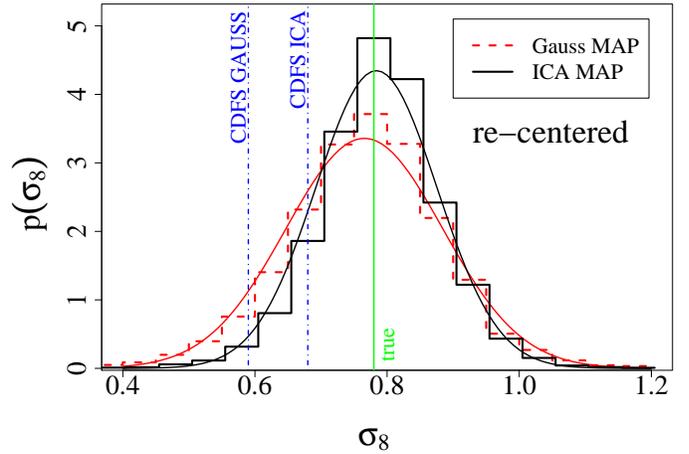}}
\caption{Same as Fig.\ \ref{fig:cdfs_s8}, but using re-centred correlation functions}
\label{fig:cdfs_s8_f}
\end{figure}

The resulting sampling distributions of $\hat{\sigma}_8$ are shown in Figs.\ \ref{fig:cdfs_s8} (original $\xi$) and \ref{fig:cdfs_s8_f} (re-centred $\xi$). All the distributions are well fit by a Gaussian. With the original correlation functions, we obtain estimates $\hat{\sigma}_8$ which are too high on average. This reflects the fact that the power spectrum fitting formula by \citet{smithps} underpredicts the small scale power in the simulations \citep[see also][]{mrrt}. If we correct for this, we see that the maximum of the ICA likelihood is a nearly unbiased estimator of $\sigma_8$ in the one-dimensional case considered here, and in addition has a lower variance than the maximum of the Gaussian likelihood.

We estimate the probability of obtaining a power spectrum normalisation as low as the one measured in the CDFS or lower, \mbox{${\rm Prob}(\hat{\sigma}_8<\hat{\sigma}_8^{\rm CDFS})$}, by the ratio of the number of realisations which fulfil this condition to the total number of simulations. These estimates agree very well with those computed from the best fitting Gaussian distribution. The results for the MAP and median estimators are summarised in Tab.\ \ref{tab:cdfs}. As expected from the above considerations, we find higher probabilities for the re-centred correlation functions. In this case, the ICA-MAP estimator yields $13\%$ for the probability of obtaining an equally low or lower $\sigma_8$ than the CDFS. This reduces to $\approx 5\%$ when the uncorrected correlation functions are used, because the misfit of our theoretical correlation functions to the simulations biases the $\sigma_8$-estimates high. If we assume that our simulations are a reasonable representation of the real Universe, we can expect the same bias when we perform fits to real data. Therefore, \mbox{${\rm Prob}(\hat{\sigma}_8<\hat{\sigma}_8^{\rm CDFS})\approx 0.05$} as derived from the uncorrected correlation functions is most likely closest to reality. The probabilities computed from the Gauss-MAP estimates are generally smaller than the ICA-MAP values because of the lower value of $\hat{\sigma}_8^{\rm CDFS}$ found using these estimators, even though the sampling distributions of the Gauss estimators are broader.

\begin{table}
	\caption{${\rm Prob}(\sigma_8<\hat{\sigma}_8^{\rm CDFS})$ for the CDFS}
	\label{tab:cdfs}
	\centering
	\begin{tabular}{c c c c c}
		\hline\hline
		& Gauss  & Gauss & ICA & ICA \\
		& (MAP) & (median) & (MAP) & (median) \\
		\hline
		re-centred CF & $6.8\%$& $8.6\%$& $12.9\%$& $9.0\%$\\
		original CF & $1.8\%$& $3.0\%$& $5.4\%$& $3.4\%$\\
		\hline
	\end{tabular}

\end{table}

\begin{figure}
	\resizebox{\hsize}{!}{\includegraphics[angle=0]{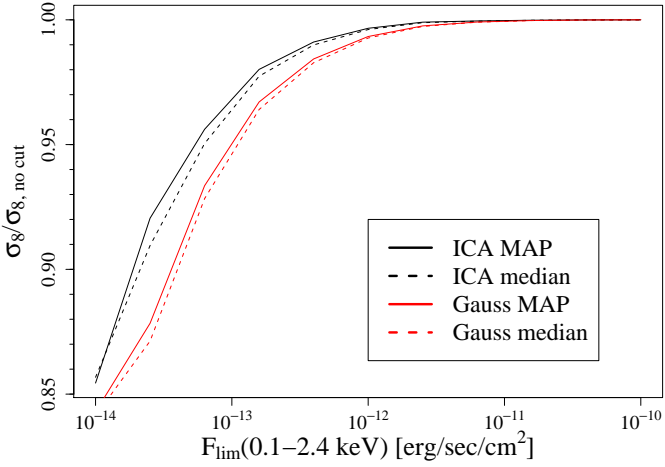}}
	\caption{The average values of the ICA-MAP (solid black line) and Gauss-MAP (solid red line) estimators computed from CDFS realisations that do not contain clusters with an X-ray flux larger than $F_{\rm lim}$. For comparison, we also plot the averages of the corresponding median estimators (dashed lines).}
	\label{fig:s8flux}
\end{figure}

\begin{figure}
	\resizebox{\hsize}{!}{\includegraphics[angle=0]{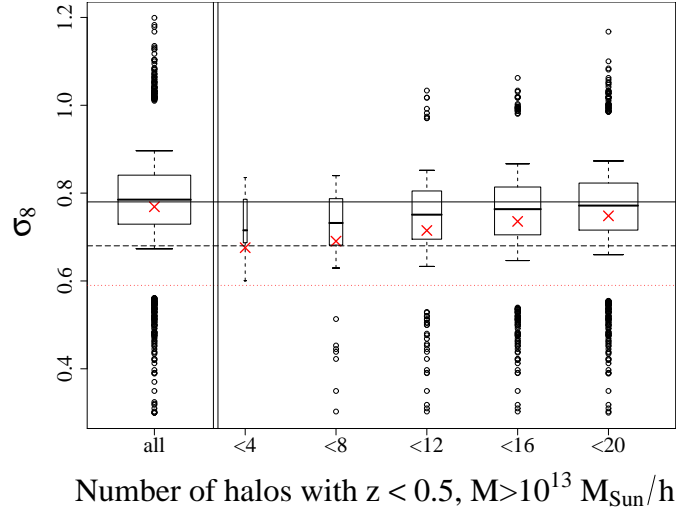}}
	\caption{Dependence of the ICA-MAP-estimator for $\sigma_8$ on the number of group- and cluster-sized haloes $n_{\rm halo}$ between $z=0$ and $z=0.5$. For each $n_{\rm halo}$-bin, we summarise the distribution of the corresponding subsample of simulated CDFS-fields by giving a box plot: the thick horizontal line in each box denotes the median, the upper and lower box boundaries give the upper and lower quartiles of the distribution of the sample values. The error bars (``whiskers'') extend to the $10\%$ and $90\%$ quantiles, respectively. To visualise the tails of the distributions, the most extreme values are given as points. The width of each box is proportional to the square root of the sample size. For comparison, we also show for each subsample the median of the Gauss MAP estimators as red crosses.	The solid black horizontal line indicates the true value of $\sigma_8$, the black dashed line the ICA-MAP estimate for the CDFS and the red dotted line the Gauss-MAP estimate. The average number of haloes with $M>10^{13}\,M_\odot$ and $z\leq 0.5$ in a CDFS-like field is $\bar{n}_{\rm halo}=18.5$.}
	\label{fig:s8boxplot}
\end{figure}

\subsection{Influence of the CDFS selection criteria}
We now investigate if and by how much the way in which the CDFS was selected can bias our estimates of the power spectrum normalisation low.
Several local criteria had to be fulfilled by the future CDFS, such as a low galactic H{\sc I} density, the absence of bright stars and observability from certain observatory sites. Since these conditions do not reach beyond our galaxy, we do not expect them to affect the lensing signal by the cosmological large-scale structure.

Furthermore, the field was chosen such that no extended X-ray sources from the ROSAT All-Sky Survey (RASS), in particular galaxy clusters, are in the field of view. This is potentially important, since it is known from halo-model calculations that the cosmic shear power spectrum on intermediate and small scales is dominated by group- and cluster-sized haloes. Therefore, the exclusion of X-ray clusters might bias the selection of a suitable line of sight towards under-dense fields. On the other hand, the RASS is quite shallow and thus only contains very luminous or nearby clusters, which have a limited impact on the lensing signal due to their low number or low lensing efficiency. We quantify the importance of this criterion using the halo catalogues of our $N$-body-simulations. To each halo, we assign an X-ray luminosity in the energy range from $0.1$ to $2.4\,{\rm keV}$ using the mass-luminosity relation given in \citet{reiprich_boehringer} and convert this into X-ray flux using the halo redshift. We then compute the average of the $\sigma_8$ estimates from all fields which do not contain a cluster brighter than a certain flux limit. It is difficult to define an exact overall flux limit to describe the CDFS selection, because the RASS is rather heterogeneous. However, it is apparent from Fig.~\ref{fig:s8flux} that even a very conservative limit of $10^{-13}\,{\rm ergs}/{\rm sec}/{\rm cm}^2$ will change the average $\sigma_8$ estimate by at most $3-5\%$. This bias is therefore most likely not large enough to explain our CDFS result alone.

Finally, the CDFS candidate should not contain any ``relevant NED source''. This is very hard to translate into a quantitative criterion, in particular because our simulations contain only dark matter.  We model the effect of imposing this requirement by demanding that there be less than $n_{\rm halo}$ group- or cluster-sized haloes ($M>10^{13}\,M_\odot/h$) in the redshift range from $z=0$ and $z=0.5$ in a CDFS candidate. The impact of this criterion on the estimated value of $\hat{\sigma}_8$ using the ICA- and Gauss-MAP  estimators is shown in Fig.\ \ref{fig:s8boxplot}. As expected, the median $\hat{\sigma}_8$ is a monotonically increasing function of $n_{\rm halo}$. For fields with less than $\approx 12$ massive haloes, the probability of obtaining a power spectrum normalisation as low as in the CDFS rises above $\approx 20\%$. Given that the average number of massive haloes in the specified redshift range is $18.5$, it does not seem to be too unreasonable that fields with less than $\approx 12$ such haloes could be obtained by selecting ``empty'' regions in the NED.
This is also in qualitative agreement with \citet{phleps07}, who find that the CDFS is underdense by a factor of $\approx 2$ in the redshift range from $z\approx 0.2$ to $z\approx 0.4$.

\begin{figure}
	\resizebox{\hsize}{!}{\includegraphics[angle=270]{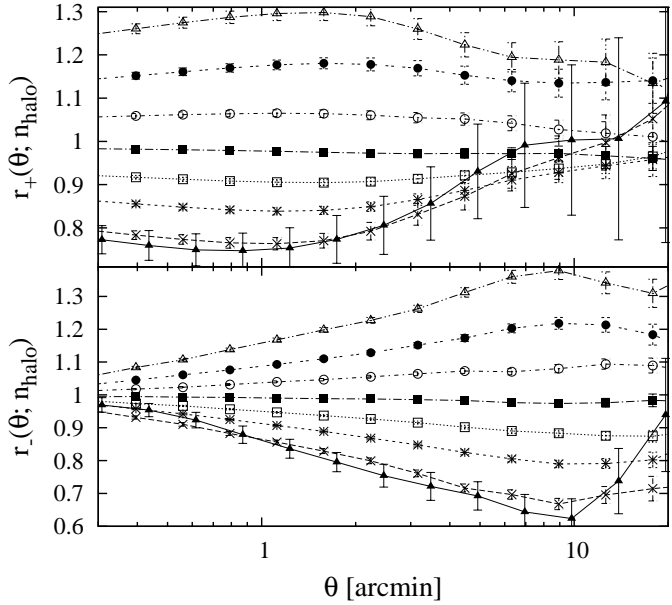}}
	\caption{Ratios $r_+$ (upper panel) and $r_-$ (lower panel) of the shear correlation functions in a particular $n_{\rm halo}$-bin to the average correlation function of all realisations. The lowest (solid) curve represents the bin with $n_{\rm halo}\in[0,4)$, the second lowest the bin with $n_{\rm halo}\in[4,8)$, and so on. The highest ratio corresponds to the bin with $n_{\rm halo}\geq 28$. The error bars have been estimated from the field-to-field variation.}
	\label{fig:cdfs_cfnuisance}
\end{figure}

\begin{figure}
	\resizebox{\hsize}{!}{\includegraphics[angle=270]{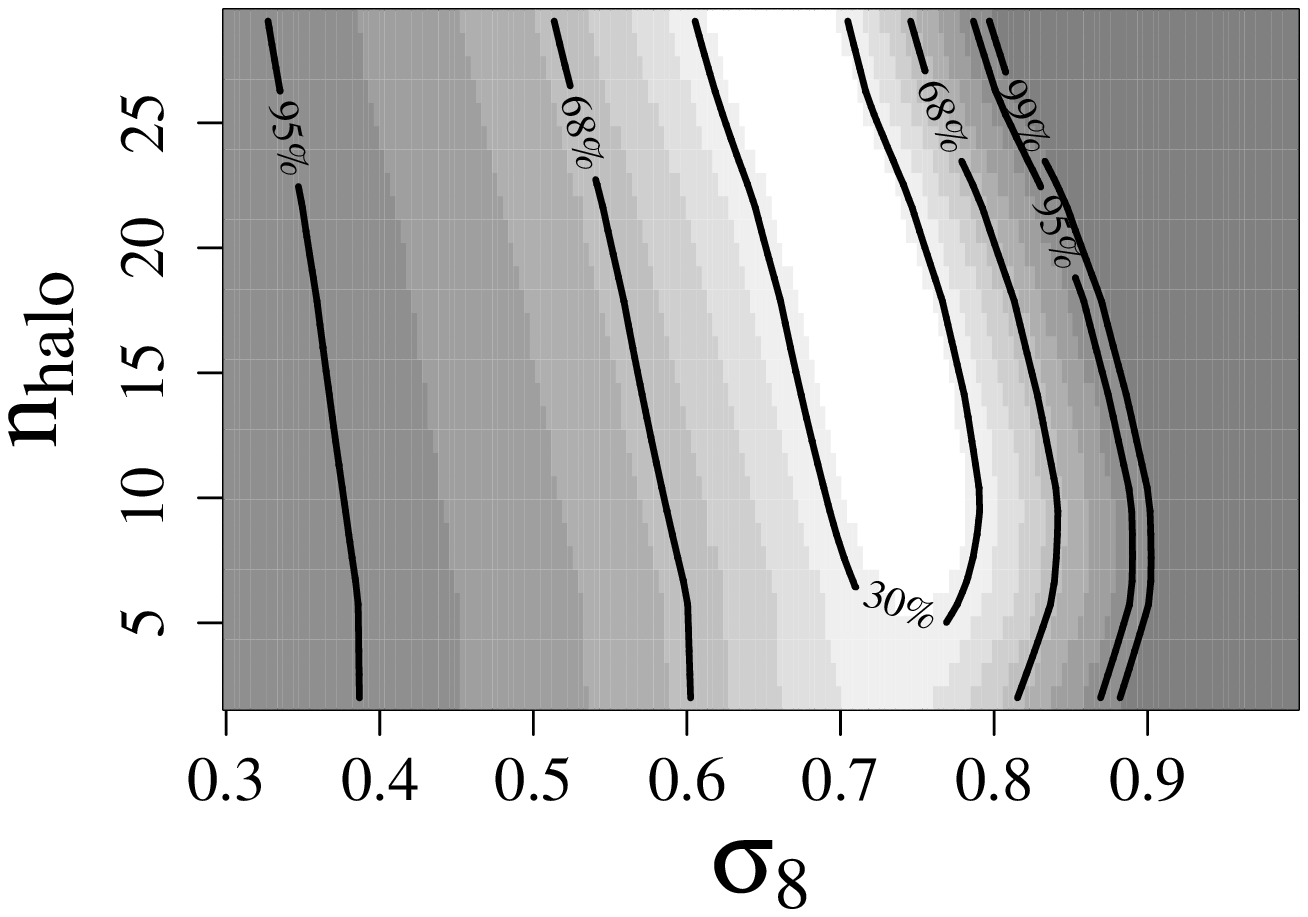}}
	\resizebox{\hsize}{!}{\includegraphics[angle=0]{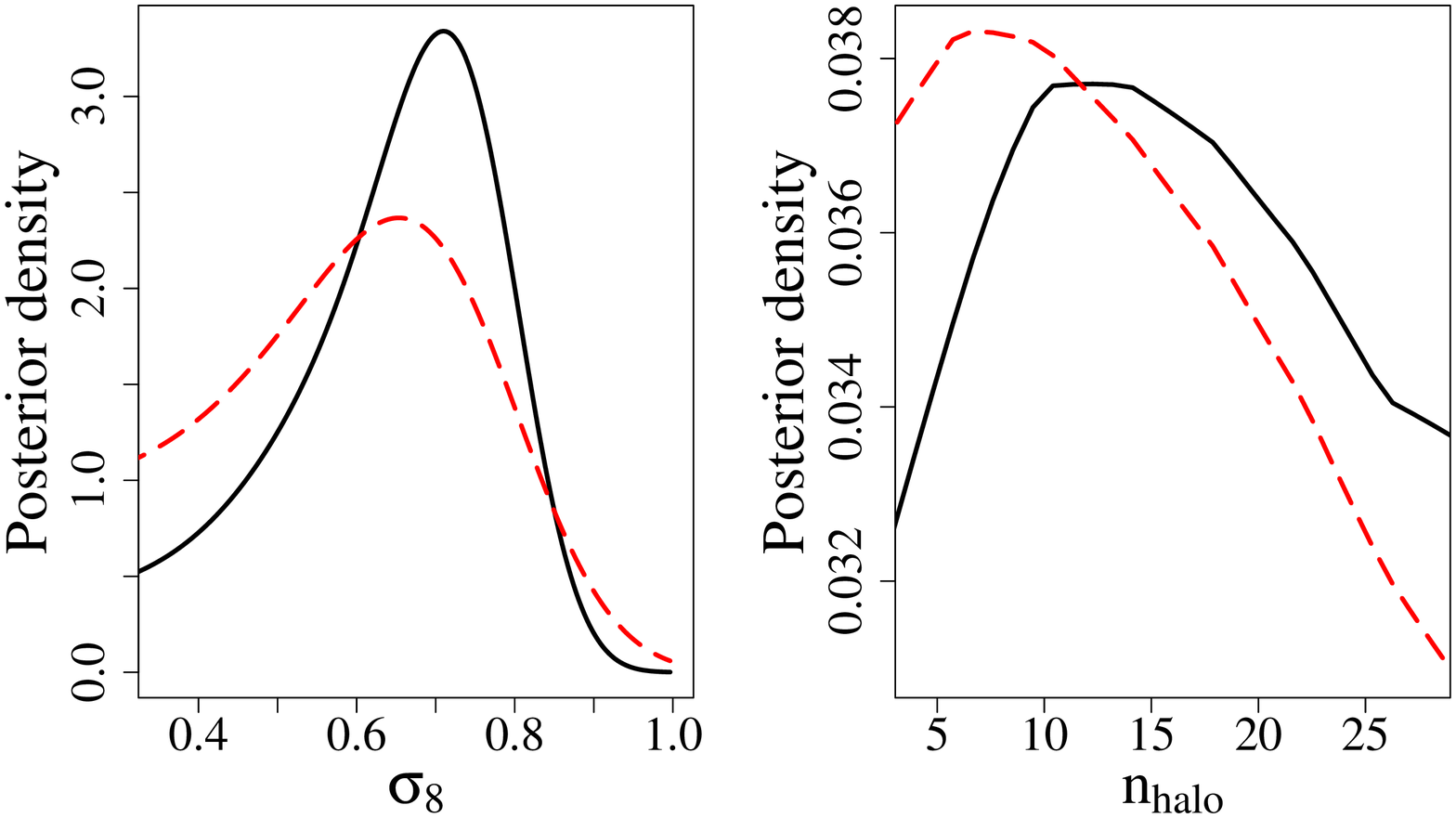}}
	\caption{Upper panel: Posterior density for $\sigma_8(\Omega_{\rm m}=0.25)$ and $n_{\rm halo}$ computed using the ICA likelihood, keeping all other cosmological parameters constant.  Lower panels: Marginalised posterior densities of $\sigma_8(\Omega_{\rm m}=0.25)$ (left panel) and $n_{\rm halo}$ (right panel). Solid black curves show the results from using the ICA likelihood, dashed red lines from the Gaussian likelihood and the ray-tracing covariance.}
	\label{fig:cdfs_s8_halocorr}
\end{figure}

We estimate the impact of this selection criterion on the estimates of cosmological parameters by treating the number of haloes in the CDFS as a nuisance parameter in the process of parameter estimation. Similar to what we did to obtain Fig.\ \ref{fig:s8boxplot}, we bin the realisations of the CDFS according to the number of group-sized haloes in the realisations. For each bin, we obtain the mean shear correlation function and its ratio to the mean shear correlation function of all realisations, $r_\pm(\theta, n_{\rm halo}) = \xi_\pm(\theta,n_{\rm halo}) / \xi_\pm(\theta)$. The functions $r_+$ and $r_-$ are shown in Fig.~\ref{fig:cdfs_cfnuisance}. The realisations with fewer (more) haloes than the average generally display a smaller (larger) shear correlation function. We fit the ratios in each bin with a double power law of the form 
\begin{equation} \label{eq:cfratio}
r_\pm(\theta,n_{\rm halo}) = A_\pm(n_{\rm halo}) \theta^{\alpha_\pm(n_{\rm halo})} + B_\pm(n_{\rm halo}) \theta^{\beta_\pm(n_{\rm halo})}\;.
\end{equation}
For values of $n_{\rm halo}$ which do not coincide with one of the bin centres, the functions $r_\pm$ are obtained by linear interpolation between the fits for the two adjacent bins. With this, we extend our model for the shear correlation function to \mbox{$m'_\pm(\theta;\vec \pi, n_{\rm halo}) = m_\pm(\theta;\vec \pi)\,r_\pm(\theta, n_{\rm halo})$}. In Fig.~\ref{fig:cdfs_s8_halocorr}, we show the resulting posterior distributions for $\sigma_8(\Omega_{\rm m}=0.25)$ and $n_{\rm halo}$, keeping all other cosmological parameters fixed and using a uniform prior for $n_{\rm halo}$. The two-dimensional distribution shows a weak correlation between the two parameters: as expected, a low (high) value of $n_{\rm halo}$ requires a slightly higher (lower) value of $\sigma_8$. The marginalised posterior for $\sigma_8$ is very similar to the one shown in Fig.~\ref{fig:cdfs_s8}, where the field selection is not taken into account. However, including $n_{\rm halo}$ increases the MAP estimate of $\sigma_8$ by $5\%$ to $\hat\sigma_8  = 0.71^{+0.10}_{-0.15}$ for the ICA likelihood and by $10\%$ to $\hat\sigma_8  = 0.65^{+0.13}_{-0.20}$ for the Gaussian likelihood and the ray-tracing covariance matrix. The marginalised posterior distribution of $n_{\rm halo}$ shows a weak peak at $\hat n_{\rm halo} \approx 13$ (compared to the average of $\bar n_{\rm halo}=18.5$ for all ray-tracing realisations) in the ICA case and even lower values if the Gaussian likelihood is used. Overall, however, the posterior is very shallow.

Having corrected for the field selection, we can now recompute the probabilities given in Tab.~\ref{tab:cdfs} for drawing the CDFS at random. We find for the ICA-MAP estimate \mbox{${\rm Prob}(\sigma_8<0.71) = 9.4\%$} for the original shear correlation functions and \mbox{${\rm Prob}(\sigma_8<0.71) = 18.5\%$} for the re-centred ones. For the Gaussian likelihood, we find \mbox{${\rm Prob}(\sigma_8<0.65) = 6.0\%$} (original) and \mbox{${\rm Prob}(\sigma_8<0.65) = 14.9\%$} (re-centred), respectively.

\begin{figure}
	\resizebox{\hsize}{!}{\includegraphics[angle=0]{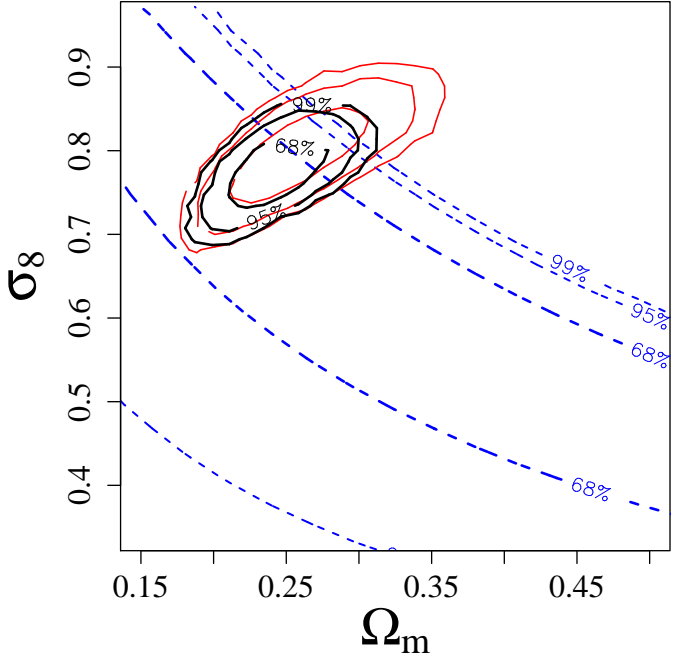}}
	\caption{Posterior density for $\Omega_{\rm m}$ and $\sigma_8$, where we have marginalised over the Hubble constant $h_{100}$ and the number of haloes in the field $n_{\rm halo}$. The dashed blue contours show the  $68\%$, $95\%$ and $99\%$ credible regions resulting from the cosmic shear analysis of the CDFS (using the ICA likelihood), the red contours show the posterior from the WMAP 5-year data (using the flat $\Lambda$CDM model). The combined posterior is shown with thick black contours. }
	\label{fig:cdfs_wmap_halocorr}
\end{figure}

With this (approximate) treatment of the systematic effects caused by the field selection, we can now put the CDFS in context with the results from the WMAP five-year data. For this, we fit the shear correlation function for $\Omega_{\rm m}$ and $\sigma_8$, marginalising over $h_{100}$ \citep[with a Gaussian prior centred on $h_{100}=0.7$ and $\sigma_{h_{100}}=0.07$, as suggested by the Hubble Key Project;][]{HSTKey} and $n_{\rm halo}$ with a uniform prior. We use the WMAP Markov chain for a flat $\Lambda$CDM model \citep[lcdm+sz+lens;][]{dunkley09,komatsu09}, where again we marginalise over all parameters except $\Omega_{\rm m}$ and $\sigma_8$. The resulting posterior distributions for the CDFS only (blue dashed contours), WMAP only (red contours) and the combination of both measurements (thick black contours) are shown in Fig.~\ref{fig:cdfs_wmap_halocorr}. Clearly, the joint posterior is dominated by the WMAP data; however, the constraints from the CDFS allow us to exclude parameter combinations where both $\Omega_{\rm m}$ and $\sigma_8$ are large. We find the MAP estimates $\hat\Omega_{\rm m}  = 0.26^{+0.03}_{-0.02}$ and $\hat\sigma_8  = 0.79^{+0.04}_{-0.03}$ when marginalising over the other parameter.

Finally, note that the two criteria discussed in this section are not strictly independent. However, it is highly improbable that a single field will contain more than one massive halo above the X-ray flux limit. Therefore, selecting fields without an X-ray-bright cluster prior to performing the steps that lead to Fig.~\ref{fig:s8boxplot} would change the halo numbers that go into the analysis by at most one and would not significantly influence the foregoing discussion.

\section{Summary and discussion}
In this paper, we have investigated the validity of the approximation of a Gaussian likelihood for the cosmic shear correlation function, which is routinely made in weak lensing studies. We have described a method to estimate the likelihood from a large set of ray-tracing simulations. The algorithm tries to find a new set of (non-orthogonal) basis vectors with respect to which the components of the shear correlation functions become approximately statistically independent. This then allows us to estimate the high-dimensional likelihood as a product of one-dimensional probability distributions. A drawback of this method is that quite a large sample of realistically simulated correlation functions is required to get good results for the tails of the likelihood. However, this should become less problematic in the near future when increasingly large ray-tracing simulations will become available.

We have investigated how the constraints on matter and vacuum energy density, Hubble parameter and power spectrum normalisation depend on the shape of the likelihood for a survey composed of $0.5\deg\times 0.5\deg$ fields and a redshift distribution similar to the CDFS. We find that if the non-Gaussianity of the likelihood is taken into account, the posterior likelihood becomes more sharply peaked and skewed. When fitting only for $\Omega_{\rm m}$ and $\sigma_8$, the maximum of the posterior is shifted towards lower $\Omega_{\rm m}$ and higher $\sigma_8$, and the area of the $68\%$ highest posterior density credible region decreases by about $40\%$ compared to the case of a Gaussian likelihood. For the four-dimensional parameter space, we have conducted a Fisher matrix analysis to obtain lower limits on the errors achievable with a $1500\,{\rm deg}^2$ survey. As in the two-dimensional case, we find the most important effect to be that the error bars decrease by $10-40\%$ compared to the Gaussian likelihood. Less severe is the slight tilt of the Fisher ellipses when marginalising over two of the four parameters, particularly when $h_{100}$ is involved.

In the second part of this work, we have presented a re-analysis of the CDFS-HST data.
Using the non-Gaussian likelihood, we find $\sigma_8=0.68_{-0.16}^{+0.09}$ for $\Omega_{\rm m}=0.25$ (keeping all other parameters fixed to their fiducial values), compared to $\sigma_8=0.59_{-0.19}^{+0.10}$ obtained from the Gaussian likelihood with a covariance matrix estimated from the ray-tracing simulations. We have then tried to quantify how (un-)likely it is to randomly select a field with the characteristics of the \textit{Chandra} Deep Field South with a power spectrum normalisation this low. We have used $9600$ ray-tracing realisations of the CDFS to estimate the sampling distribution of the ICA-MAP estimator for $\sigma_8$. For our fiducial, WMAP5-like cosmology, we find that ${\rm Prob}(\sigma_8\leq 0.68) \approx 5\%$, assuming that the location of the CDFS on the sky was chosen randomly. The fact that the CDFS was selected not to contain an extended X-ray source in the ROSAT All-Sky Survey can lead to a bias of the estimated $\sigma_8$ by at most $5\%$. This is because the clusters excluded by this criterion are rare and mostly at low redshifts, and therefore not very lensing-efficient. The second relevant selection criterion is that the CDFS should not contain any relevant NED source. We model this by selecting only those fields which contain a specific number $n_{\rm halo}$ of group- and cluster-sized haloes. We find that for those realisations for which the number of such haloes is below the average, the estimates of $\sigma_8$ can be biased low by about \mbox{$5$-$10\%$}. We include this effect in our likelihood analysis by extending our model shear correlation function by a correction factor depending on $n_{\rm halo}$ and treating $n_{\rm halo}$ as a nuisance parameter. This increases the estimate of $\sigma_8$ by $5\%$ to $\hat\sigma_8  = 0.71^{+0.10}_{-0.15}$ for the ICA likelihood and by $10\%$ to $\hat\sigma_8  = 0.65^{+0.13}_{-0.20}$ for the Gaussian likelihood. This procedure also yields tentative evidence that the number of massive haloes in the CDFS is only $\approx 70\%$ of the average, in qualitative agreement with the findings of \citet{phleps07}.

Finally, we combine the CDFS cosmic shear results with the constraints on cosmological parameters from the WMAP experiment. We fit for $\Omega_{\rm m}$ and $\sigma_8$, where  we marginalise over the Hubble constant and take into account the field selection bias by marginalising also over $n_{\rm halo}$. While the posterior is clearly dominated by the WMAP data, the CDFS still allows us to exclude parts of the parameter space with high values of both $\Omega_{\rm m}$ and $\sigma_8$. Assuming a flat Universe, the MAP estimates for these two parameters are $\hat\Omega_{\rm m}  = 0.26^{+0.03}_{-0.02}$ and $\hat\sigma_8  = 0.79^{+0.04}_{-0.03}$.

\begin{acknowledgements}
We would like to thank Tim Eifler for useful discussions and the computation of the Gaussian covariance matrix, and Sherry Suyu for careful reading of the manuscript. We thank Richard Massey and William High for providing the STEP2 image simulations.
JH acknowledges support by the Deutsche Forschungsgemeinschaft within the Priority Programme 1177 under the project SCHN 342/6 and by the Bonn-Cologne Graduate School of Physics and Astronomy. TS acknowledges financial support from the Netherlands Organization for Scientific Research (NWO) and the German Federal Ministry of Education and Research (BMBF) through the TR33 ``The Dark Universe''. We acknowledge the use of the Legacy Archive for Microwave Background Data Analysis (LAMBDA). Support for LAMBDA is provided by the NASA Office of Space Science.
\end{acknowledgements}


\appendix

\section{Projection Pursuit Density Estimation}\label{sec:ppde}

In order to have an independent check of the ICA-based likelihood estimation algorithm, we employ the method of \emph{projection pursuit density estimation} \citep[PPDE;][]{PPDE}. Like our ICA method, PPDE aims to estimate the joint probability density $p(\vec x)$ of a random vector $\vec x$, given a set of observations of $\vec x$. As starting point, an initial model $p_0(\vec x)$ for the multidimensional probability distribution $p(\vec x)$ has to be provided, for which a reasonable choice is e.g. a multivariate Gaussian with a covariance matrix estimated from the data. The method then identifies the direction $\vec \theta_1$ along which the marginalised model distribution differs most from the marginalised density of the data points and corrects for the discrepancy along the direction $\vec\theta_1$ by multiplying $p_0$ with a correction factor. This yields a refined density estimate $p_1(\vec x)$, which can be further improved by iteratively applying the outlined procedure.

More formally, the PPDE density estimate is of the form
\begin{equation} \label{eq:stat_ppde1}
	p_M(\vec x) = p_0(\vec x)\,\prod_{m=1}^M\,f_m(\vec\theta_m\cdot \vec x)\;,
\end{equation}
where $p_M$ is the estimate after $M$ iterations of the procedure and $p_0$ is the initial model. The univariate functions $f_m$ are multiplicative corrections to the initial model along the directions $\vec\theta_m$. The density estimate can be obtained iteratively using the relation $p_M(\vec x) = p_{M-1}(\vec x)\,f_M(\vec\theta_M\cdot \vec x)$. At the $M$-th step of the iteration, a direction $\vec\theta_M$ and a function $f_M$ are chosen to minimise the Kullback-Leibler divergence \citep{KLdiv} between the actual data density $p(\vec x)$ and the density estimate $p_M(\vec x)$,
\begin{equation}
	D_{\rm KL}[p,p_M] = \int\dd\vec x\;p(\vec x)\,\log \frac{p(\vec x)}{p_M(\vec x)}\;,
\end{equation}
as a goodness-of-fit measure.
The Kullback-Leibler divergence provides a ``distance measure'' between two probability distribution functions, since it is non-negative and zero only if \mbox{$p\equiv q$}, albeit not symmetric.
Only the cross term
\begin{equation}
	W(\vec\theta_M,f_M)= -\int\dd\vec x\;p(\vec x)\,\log p_M(\vec x)
\end{equation}
of the K-L divergence is relevant for the minimisation, all other terms do not depend on $\vec\theta_M$ and $f_M$. By using Eq.~\eqref{eq:stat_ppde1}, one sees that the minimum of $W$ is attained at the same location as the minimum of
\begin{equation} \label{eq:stat_ppde2}
	w(\vec\theta_M,f_M)=-\int\dd\vec x\;p(\vec x)\,\log f_M(\vec\theta_M\cdot\vec x)\;,
\end{equation}
which is the expectation value of $\log f_M$ with respect to $p(\vec x)$. The data density $p(\vec x)$ is unknown; however, the data comprise a set of $N$ samples from this distribution. The expectation value of $\log f_M$ can therefore be estimated by
\begin{equation}
	\hat w(\vec\theta_M,f_M)=-\frac{1}{N}\sum_{i=1}^N \log f_M(\vec\theta_M\cdot\vec x_i)\;.
\end{equation}
For fixed $\vec\theta_M$, the minimum of Eq.~\eqref{eq:stat_ppde2} is attained for
\begin{equation}\label{eq:stat_ppde3}
	f_M(\vec\theta_M\cdot\vec x) =  \frac{p^{\vec\theta_M}(\vec\theta_M\cdot\vec x)}{p^{\vec\theta_M}_{M-1}(\vec\theta_M\cdot\vec x)}\;,
\end{equation}
where $p^{\vec\theta_M}$ and $p^{\vec\theta_M}_{M-1}$ are the marginal densities of the data and of model density from the $(M-1)$-st iteration along the direction $\vec\theta_M$, respectively. With this, the iterative process that leads to estimates of $\vec\theta_M$ and $f_M$ schematically consists of:
 \begin{itemize}
 	\item choosing a direction $\vec\theta_M$,
 	\item computing the marginal densities $p^{\vec\theta_M}$ and $p^{\vec\theta_M}_{M-1}$,
 	\item computing $f_M(\vec\theta_M\cdot\vec x)$ according to Eq.~\eqref{eq:stat_ppde3},
 	\item computing $\hat w(\vec\theta_M,f_M)$
 	\item choosing a new $\vec\theta_M$ that decreases $\hat w$
 	\item continuing from step 2 until a convergence criterion is fulfilled.
 \end{itemize}
To efficiently compute the marginals $p^{\vec\theta_M}$ and $p^{\vec\theta_M}_{M-1}$, Monte Carlo samples of these densities are used. Note that the data already comprise a sample of $p(\vec x)$; a sample of $p^{\vec\theta_M}_{M-1}$ can be obtained efficiently by an iterative method: since $p_{M-1}$ is similar to $p_{M-2}$, a subset of the sample from $p_{M-1}$ can be obtained by rejection sampling from the sample from the $(M-2)$-nd step. The remaining data vectors are then drawn by rejection sampling from $p_0$. For more technical details of the estimation procedure, we refer the reader to \citet{PPDE}.

Note that the PPDE technique, although using very similar methodology as our ICA-based procedure, is different in the important point that it does not rely on the assumption that a linear transformation of the data leads to statistical independence of the components of the transformed data vectors. It therefore comprises a good test of the validity of this approximation.

\section{Fisher matrix of the ICA likelihood}\label{sec:icafisher}


In this appendix, we give the derivation of Eq.~\eqref{eq:icafisher_deriv}. In
the general case, the Fisher matrix is given by \citep[e.g.][]{kendall_stuart}
\begin{equation} \label{eq:fisherdef_app}
	\tens{F}_{\alpha\beta} = \left\langle \frac{\partial \log L}{\partial \pi_\alpha} \frac{\partial \log L}{\partial \pi_\beta} \right\rangle \;\; .
\end{equation}
In our case, the likelihood depends on
cosmological parameters only through the difference between data and model
vector, i.e. $\vec{s}=\breve{\vec{\xi}} - \breve{\vec{m}}$ (see Eq.~\ref{eq:sdef}).
This allows us to write
\begin{eqnarray}
	\frac{\partial \log L(\vec{s}(\vec{\pi}))}{\partial \pi_\alpha} &=& \frac{\partial \log L(\vec{s})}{\partial s_i} \; \frac{\partial s_i}{\partial \pi_\alpha}\\
	&=& \frac{\dd \log p_{s_i}(s_i)}{\dd s_i} \; \frac{\partial s_i}{\partial \pi_\alpha}\;\;,
\end{eqnarray}
where in the last step we have made use of the fact that the likelihood factorises in the ICA basis.
The expression for the Fisher matrix then can be written as
\begin{equation}
	\tens{F}_{\alpha\beta} =\sum_{i,j}\,\left\langle \frac{\dd \log p_{s_i}(s_i)}{\dd s_i}  \frac{\dd \log p_{s_j}(s_j)}{\dd s_j} \right\rangle \; \frac{\partial \breve{m}_i}{\partial \pi_\alpha} \frac{\partial \breve{m}_j}{\partial \pi_\beta}
\end{equation}
Next, we compute the expectation value on the right hand side and obtain
\begin{eqnarray}
	\tens{F}_{\alpha\beta} &=& \sum_{i\neq j}\;  \frac{\partial \breve{m}_i}{\partial \pi_\alpha} \frac{\partial \breve{m}_j}{\partial \pi_\beta}\; \int\dd s_i\; \frac{\dd p_{s_i}(s_i)}{\dd s_i} \int\dd s_j \frac{\dd p_{s_j}(s_j)}{\dd s_j}\\
	&&+ \sum_{i}\; \frac{\partial \breve{m}_i}{\partial \pi_\alpha} \frac{\partial \breve{m}_i}{\partial \pi_\beta}\; \int\dd s_i\; p_{s_i}(s_i)\,\left(\frac{\dd \log p_{s_i}(s_i)}{\dd s_i}\right)^2 \;  \;\;.
\end{eqnarray}
The integrals in the first term of the right hand side vanish since the $p_{s_i}$ drop to zero for very large and small values of $s_i$. This leaves us with
\begin{equation} \label{eq:icafisher_app}
	\tens{F}_{\alpha\beta} = \sum_{i}\; \frac{\partial \breve{m}_i}{\partial \pi_\alpha} \frac{\partial \breve{m}_i}{\partial \pi_\beta}\; \int\dd s_i\; p_{s_i}(s_i)\,\left(\frac{\dd \log p_{s_i}(s_i)}{\dd s_i}\right)^2 \;\;.
\end{equation}

The derivatives in Eq.~\eqref{eq:icafisher_app} can be strongly affected by noise in the estimated $p_{s_i}(s_i)$, in particular in the tails of the distributions. For their numerical computation, we therefore choose the following four-point finite difference operator \citep{abramowitz_stegun}:
\begin{equation}
	\frac{\dd p}{\dd s} = \frac{p(s-2h)-8p(s-h)+8p(s+h)-p(s+2h))}{12h} + {\rm O}(h^5)\;\;,
\end{equation}
which we find to be more stable against this problem than its more commonly used two-point counterpart.
Because of this potential difficulty, we cross-check our results with the alternative method provided by Eq.~\eqref{eq:icafisher_exp}. This method is significantly slower, but numerically simpler. This is because the derivatives of the log-likelihood in Eq.~\eqref{eq:icafisher_exp} are on average computed close to the maximum-likelihood point, where the likelihood estimate is well sampled.  Reassuringly, we find excellent agreement between the two methods. Finally, we have investigated the influence of the choice of the Kernel function $K$ in Eq.~\eqref{eq:kerndensest}, which might affect the computation of the numerical derivatives. Our results prove to be stable against variation of $K$, provided that we chose a differentiable Kernel function.

\section{Further conclusions for our KSB+ pipeline from the STEP simulations}
\label{se:ap:step2}
In this appendix we assume that the reader is familiar with basic KSB
notation. For a short introduction and a summary of differences between
various implementations see  \citet{hwb06}.

Within the Shear TEsting
Programme\footnote{\url{http://www.physics.ubc.ca/~heymans/step.html}}
(STEP) simulated images containing sheared galaxies are analysed in blind
tests, in order to test the shear measurement accuracy of weak lensing
pipelines.
In these analyses the shear recovery accuracy has been quantified in terms
of a multiplicative calibration bias $m$ and additive PSF residuals $c$.
From the analysis of the first set of simulations \citep[STEP1,][]{hwb06},
which mimic simplified ground-based observations, we
find that our KSB+ implementation significantly under-estimates
gravitational shear on average if no calibration correction is applied.
After the elimination of selection and weighting-related effects this shear
calibration bias amounts to a relative under-estimation of $m=-9\%$.
According to our testing the largest contribution to this bias originates
from the inversion of the $P^g$-tensor, which describes the response of
galaxy
ellipticity to gravitational shear. While a full-tensor inversion reduces
this bias, it strongly increases the measurement noise \citep[see also
][]{ewb01} and dependence on galaxy selection criteria.
We therefore decided to  stick to the commonly applied approximation of
$(P^g)^{-1}=2/\mathrm{Tr}[P^g]$, which we measure from individual galaxies, and correct the shear estimate using a
multiplicative calibration factor of $c_\mathrm{cal}=1.10$ in the \citetalias{schrabbaACS} analysis.
This average calibration correction was found to be stable to the $\sim 2\%$-level between
different STEP1 simulation subsets. However, note that the bias depends on the
details of the KSB implementation, which might explain some of the scatter
between the results for different KSB codes in STEP1.
In particular, we identified a strong dependence on the choice of the
Gaussian filter scale $r_\mathrm{g}$, which is used in the computation of
the KSB brightness moments. For example changing from our default
$r_\mathrm{g}=1.0 \thinspace r_\mathrm{f}$, where $r_\mathrm{f}$ is the flux
radius as measured by SExtractor
\citep{bea96},
to $r_\mathrm{g}=0.7\thinspace  r_\mathrm{f}$, worsens the bias to $m=-17\%$.

The average calibration correction likewise proved to be robust for the second set of
image simulations \citep[STEP2,][]{mhb07}, which also mimics ground-based
data but takes into account
more complex PSFs and galaxy morphologies by applying the shapelets
formalism \citep{mrc04}.
Yet, the  STEP2 analysis revealed a significant magnitude dependence of the shear
recovery accuracy for our implementation, with a strong deterioration at faint magnitudes.
In this analysis we applied the same signal-to-noise cut \mbox{$\mathrm{S}/\mathrm{N}>4.0$} as in STEP1 \citep[KSB $\mathrm{S}/\mathrm{N}$ as defined in][]{ewb01},
where we however ignored the strong noise correlations present in the
STEP2 data, which was added to mimic the influence of drizzle.

\begin{figure}
	\centering
	\includegraphics[width=6.cm]{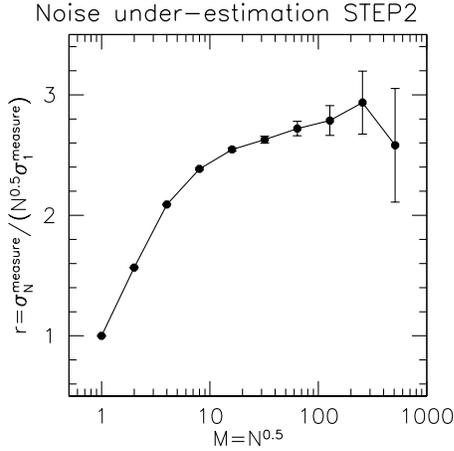}
\caption{Estimate of the effective influence of the noise correlations in
	the STEP2 simulations: Plotted is the ratio of the pixel value dispersion $\sigma_N^\mathrm{measure}$ measured from large
	areas of $N=M^2$ pixels to the estimate from the normal single pixel dispersion
	$\sqrt{N} \sigma_1^\mathrm{measure}$ as a function of $M$, determined from an
	object-free STEP2 image. In the absence of
	noise correlations $r=1$ for all $M$. The value $r\simeq 2.8$ for
	$M\rightarrow \infty$ gives the factor by which the signal-to-noise
	is over-estimated when measured from the single pixel dispersion
	$\sigma_1^\mathrm{measure}$ ignoring the correlations.
}
	\label{fi:step2:noisecorrel}
	\end{figure}

In the case of uncorrelated noise the dispersion of the sum over the pixel
values of $N$ pixels
scales as
\begin{equation}
\label{eq:step:disp_npix}
\sigma_N=\sqrt{N}\sigma_1 \,,
\end{equation}
where $\sigma_1$ is the dispersion computed from single pixel values.
Drizzling, or convolution in the case of the STEP2 simulations, reduces
$\sigma_1$ but introduces
correlations between neighbouring pixels.
The signal-to-noise of an object is usually defined as the ratio of the summed
object flux convolved with some window or weight function, divided by an
rms estimate for the noise in an equal area convolved with the same weight
function. If the noise estimate is computed from $\sigma_1$ and scaled according to
Eq.\thinspace\ref{eq:step:disp_npix},
the correlations are neglected and the noise estimate is too small.

In order to estimate the effective influence
of the  noise correlations in STEP2, we use a pure noise image which was
provided together with the simulated images.
We compute the rms of the pixel sum $\sigma_N^\mathrm{measure}$ in independent quadratic
sub-regions of the image with side length $M=\sqrt{N}$ and determine the ratio
\begin{equation}
r=\frac{\sigma_N^\mathrm{measure}}{\sqrt{N}\sigma_1^\mathrm{measure}}\,,
\end{equation}
which in the absence of correlated noise would be equal to 1 for all $N$.
In the presence of noise correlations it will  for large
$N$ converge to the factor by which
$\sigma_1^\mathrm{measure}$ under-estimates the uncorrelated  $\sigma_1$.
This can be understood as drizzling or convolution typically re-distributes
pixel flux within a relatively small area. As soon as this kernel is much smaller
than the area spanned by
$M^2$ pixels, the correlations become unimportant for the area pixel sum.
The measured $r(M)$ is plotted in Fig.\thinspace\ref{fi:step2:noisecorrel}.
Extrapolating to $M\rightarrow \infty$ we estimate that ordinary noise
measures based on the single pixel dispersion, which ignore the noise
correlation, will over-estimate the signal-to-noise of objects by a factor
$r\simeq 2.8$.
Hence, our original selection criterion \mbox{$\mathrm{S}/\mathrm{N}>4.0$} corresponds to a very low true cut
$\mathrm{S}/\mathrm{N}^\mathrm{true}\gtrsim 1.4$ including much noisier
objects  than in STEP1.

	\begin{figure}
	\centering
\includegraphics[width=5cm]{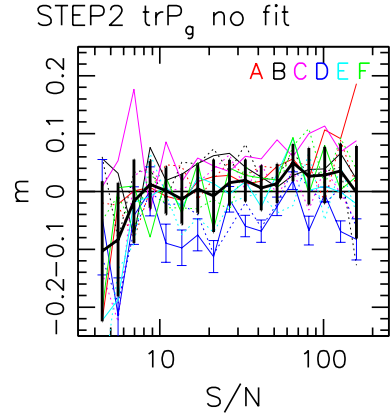}
	\caption{Calibration bias $m$ as a function of the uncorrected
		KSB signal-to-noise
		\mbox{$\mathrm{S}/\mathrm{N}$}  for the TS analysis of the STEP2
		simulations.
Thin solid (dashed) lines show $\gamma_1$ ($\gamma_2$) estimates for individual PSFs, where we show individual error-bars only for one PSF for clarity.
The bold solid line and error-bars show the mean and standard deviation of the individual PSF estimates and shear components.
Note the deterioration of the shear estimate for the STEP2 galaxies
with \mbox{$\mathrm{S}/\mathrm{N}\lesssim 7$}
(\mbox{$\mathrm{S}/\mathrm{N}^\mathrm{true}\lesssim 2.5$}).
	For this plot an adapted calibration correction of 1.08
		was applied.
}
	\label{fi:step2_sn_rh}
	\end{figure}

	\begin{figure}[htb]
	\centering
\includegraphics[width=8cm]{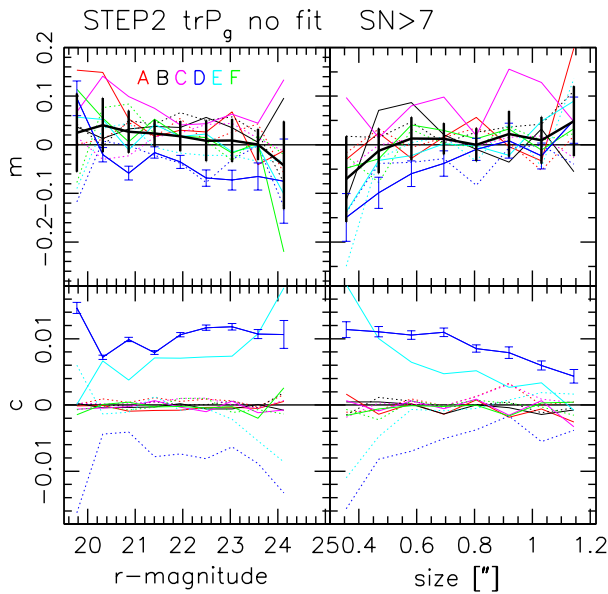}
\caption{Calibration bias $m$ and PSF residuals $c$ as a function of input
	galaxy magnitude and size for our refined analysis of the STEP2
	data.
Thin solid (dashed) lines show $\gamma_1$ ($\gamma_2$) estimates for individual PSFs, where we include individual error-bars only for one PSF for clarity.
Bold solid lines and error-bars show the mean and standard deviation of the
individual PSF estimates and shear components.
In this plot only galaxies with \mbox{$\mathrm{S}/\mathrm{N}>7$} (\mbox{$\mathrm{S}/\mathrm{N}^\mathrm{true}>2.5$})
	are taken into account, which strongly reduces the  deterioration found
	in \citet{mhb07} for faint magnitudes.
	For this plot an adapted calibration correction of 1.08
		was applied.
}
	\label{fi:step2_magsize}
	\end{figure}

We plot the dependence of our STEP2 shear estimate on the (uncorrected)
$\mathrm{S}/\mathrm{N}$
in Fig.\thinspace\ref{fi:step2_sn_rh}.
For $\mathrm{S}/\mathrm{N}\lesssim 7$, corresponding to $\mathrm{S}/\mathrm{N}^\mathrm{true}\lesssim 2.5$, a significant
deterioration of the shear signal occurs, with a mean calibration bias \mbox{$\langle
m\rangle\sim -10\%$} and a large scatter between the different PSF models.
We conclude that this approximately marks the limit down to which our KSB+ implementation
can reliably measure shear.
If we apply a modified cut $\mathrm{S}/\mathrm{N}>7.0$ to the STEP2
galaxies, the resulting  magnitude and size dependence of the shear calibration bias
is \mbox{$\lesssim\pm 5\%$} (top panels in Fig.\thinspace\ref{fi:step2_magsize}).
The remaining galaxies are best corrected with a slightly
reduced calibration factor \mbox{$c_\mathrm{cal}=1.08$}, which we apply for
the plots shown in this appendix and the updated shear analysis presented in
this paper.
The
difference between the
calibration corrections derived from STEP1 and STEP2 agrees with the estimated $\sim 2\%$
accuracy.
Note that the error increases for the highly elliptical PSFs D and E
(\mbox{$e^*\simeq 12\%$}) in STEP2, for which in addition significant PSF
anisotropy residuals occur (bottom panels in
Fig.\thinspace\ref{fi:step2_magsize}). This should however not affect our
analysis given that typical ACS PSF ellipticities rarely exceed
\mbox{$e^*\simeq 7\%$}, see e.g.\thinspace\citetalias{schrabbaACS}.

\bibliographystyle{aa}
\bibliography{11697}

\end{document}